\documentclass[11pt]{article}
\usepackage{jcappub,natbib,float,caption}
\title{Galaxy redshift surveys with sparse sampling}
\author[a]{Chi-Ting Chiang,}
\author[a]{Philipp Wullstein,}
\author[b]{Donghui Jeong,}
\author[a,c,d,e]{Eiichiro Komatsu,}
\author[f]{Guillermo A. Blanc,}
\author[g,h]{Robin Ciardullo,}
\author[k]{Niv Drory,}
\author[i,j]{Maximilian Fabricius,}
\author[e]{Steven Finkelstein,}
\author[d,e]{Karl Gebhardt,}
\author[g,h]{Caryl Gronwall,}
\author[g,h]{Alex Hagen,}
\author[d,e,k]{Gary J. Hill,}
\author[a]{Inh Jee,}
\author[e]{Shardha Jogee,}
\author[i]{Martin Landriau,}
\author[e]{Erin Mentuch Cooper,}
\author[g,h]{Donald P. Schneider,}
\author[e,k]{Sarah Tuttle}
\affiliation[a]{Max-Planck-Institut f\"ur Astrophysik, Karl-Schwarzschild-Str. 1, 85741 Garching, Germany}
\affiliation[b]{Department of Physics and Astronomy, Johns Hopkins
University, Baltimore, Maryland 21218, USA}
\affiliation[c]{Kavli Institute for the Physics and Mathematics of the
Universe, Todai Institutes for Advanced Study, the University of Tokyo,
Kashiwa, Japan 277-8583 (Kavli IPMU, WPI)}
\affiliation[d]{Texas Cosmology Center, The University of Texas at Austin, 2515 Speedway, Stop C1400, Austin, Texas 78712-1205, USA}
\affiliation[e]{Department of Astronomy, The University of Texas at Austin, 2515 Speedway, Stop C1400, Austin, Texas 78712-1205, USA}
\affiliation[f]{Observatories of the Carnegie Institution for Science,
813 Santa Barbara Street, Pasadena, CA 91101, USA}
\affiliation[g]{Department of Astronomy and Astrophysics, The Pennsylvania State University,
University Park, PA 16802, USA}
\affiliation[h]{Institute for Gravitation and the Cosmos, The Pennsylvania State University,
University Park, PA 16802, USA}
\affiliation[i]{Max-Planck-Institut f\"ur Extraterrestrische Physik, Postfach 1312,
Giessenbachstra\ss e, 85748 Garching, Germany}
\affiliation[j]{University Observatory Munich, Schienerstra\ss e 1, 81679, Munich, Germany}
\affiliation[k]{McDonald Observatory, The University of Texas at Austin, 2515 Speedway, Stop C1400,
Austin, Texas 78712-1205, USA}
\emailAdd{ctchiang@mpa-garching.mpg.de}
\abstract{%
Survey observations of the three-dimensional locations of galaxies are a
powerful approach to measure the distribution of matter in the universe,
which can be used to learn about the nature of dark energy,
physics of inflation, neutrino masses, etc. A competitive survey, however,
requires a large volume (e.g., $V_{\rm survey}\sim 10~{\rm Gpc}^3$) to
be covered, and thus tends to be expensive.
A ``sparse sampling'' method offers a more affordable solution to this
problem: within a survey footprint covering a given survey volume,
$V_{\rm survey}$, we observe only a fraction of the volume. The distribution
of observed regions should be chosen such that their separation is smaller than
the length scale corresponding to the wavenumber of interest. Then
one can recover the power spectrum of galaxies with precision expected for
a survey covering a volume of $V_{\rm survey}$ (rather than the volume
of the sum of observed regions) with the number density of galaxies given
by the total number of observed galaxies divided by $V_{\rm survey}$
(rather than the number density of galaxies within an observed region).
We find that regularly-spaced sampling yields an unbiased power spectrum with
no window function effect, and deviations from regularly-spaced sampling,
which are unavoidable in realistic surveys, introduce calculable window
function effects and increase the uncertainties of the recovered power spectrum.
On the other hand, we show that the two-point correlation function (pair
counting) is not affected by sparse sampling.
While we discuss the sparse sampling method within the context of the
forthcoming 
Hobby-Eberly Telescope Dark Energy Experiment, the method is general and can be
applied to other galaxy surveys.
}
\begin{document}
\maketitle
\flushbottom
\section{Introduction}
How do we measure the large-scale distribution of matter in the universe?
Broadly speaking, there are two classes of methods: (1) to measure locations
of collapsed objects (galaxies and clusters of galaxies), and (2) to measure
the distribution of the intervening matter by gravitational lensing or
absorption lines (e.g., Lyman-$\alpha$ forest). In this paper, we focus on
the first class (although our one-dimensional study may have some relevance
to the Lyman-$\alpha$ forest), and ask the question, ``What is the most
efficient way to measure three-dimensional locations of collapsed objects?''

Ideally, one aims to measure angular positions and redshifts of
all galaxies down to a certain limiting line flux or magnitude over the full
sky; however, this approach is usually too expensive to carry out in
practice. The most time-consuming aspect is the spectroscopic
observations required to determine redshifts.
A conventional method has been to execute less-expensive imaging
surveys over some fraction of the full sky, and use certain criteria
to select candidates for spectroscopic observations. Still,
in some cases too many candidates remain for feasible spectroscopic
observations. Also, for multi-object spectrographs, collisions of
fibers/slits 
make it difficult to observe crowded (over-dense) regions.
To observe all galaxies in these regions, one must visit the same area
more than once, requiring a further selection process. As one wishes to avoid
introducing artificial clustering of galaxies due to selection effects,
selection of candidates is done such that the outcome is a fair sample of the
underlying distribution of galaxies, i.e., Poisson sampling of galaxies
selected from the imaging survey data \cite{kaiser:1986b}. This method
works and has been repeatedly used in the past: recent examples include
the Sloan Digital Sky Survey (SDSS; \cite{tegmark/etal:2004,reid/etal:2010}),
the Two-degree Field Galaxy Redshift Survey (2dFGRS; \cite{cole/etal:2005}),
the WiggleZ Dark Energy Survey \cite{blake/etal:2010},
and the VIMOS Public Extragalactic Redshift Survey
(VIPERS; \cite{guzzo/etal:2013}).

There is a way to avoid pre-selection of targets. With the advent of
Integral Field Unit (IFU) spectrographs, it is now possible to obtain
blind, multiple spectra of all points in the sky simultaneously in the
two-dimensional field-of-view of the instrument without target
pre-selection. For example, 
the Visible Integral-Field Replicable Unit Spectrograph
(VIRUS; \cite{hill/etal:2010,hill/etal:2012}), which will outfit the
10-m Hobby-Eberly Telescope (HET; \cite{ramsey/etal:1998}) at McDonald
Observatory in West Texas, consists of 75 IFUs. Each IFU has 448 fibers
and feeds one unit with two spectrographs. 150 spectrographs
(33,600 fibers) are being built and will be used for a forthcoming
galaxy survey called the Hobby-Eberly Telescope Dark Energy Experiment
(HETDEX; \cite{hill/etal:2008}). HETDEX is a blind spectroscopic survey
of emission-line galaxies in a high-redshift universe. Specifically,
HETDEX will observe spectra between 3500 and 5500 \AA\ at the resolution
of $R\sim 700$ in order to detect approximately 0.8 million Lyman-$\alpha$
emitting galaxies over the redshift range of $z=1.9-3.5$. The total 
volume covered by the survey footprints is approximately 9~Gpc$^3$.

While it is in principle possible to obtain spectra (hence redshifts) of all
galaxies in the sky down to a certain limiting line flux and within a certain
redshift range by simply ``tiling the sky with IFUs,'' in practice it still
requires a great commitment of resources. 

At this point, we first must decide what we would want to accomplish with our
galaxy survey data. In this paper, we shall focus on measuring the power
spectrum of galaxies, $P_g(k)$, from three-dimensional locations of
galaxies found by a galaxy redshift survey such as HETDEX.
If the focus is to measure the power spectrum of galaxies,
the observational requirements are somewhat relaxed. The fractional statistical
r.m.s. uncertainty of the measured power spectrum is given by
$\delta P_g(k)/P_g(k)\propto V_{\rm survey}^{-1/2}(1+[n_gP_g(k)]^{-1})$,
where $n_g$ is the number density of observed galaxies and $V_{\rm survey}$
is a survey volume. The first and second terms within the parenthesis represent 
uncertainties from sample variance and shot noise, respectively.
We would not be able to reduce the uncertainty of the measured power spectrum
significantly by observing more galaxies once we have the number density
of galaxies that satisfies $n_gP_g(k)>1$.\footnote{As $P_g(k)$ is a
decreasing function of $k$ for 
$k\gtrsim 10^{-2}~h~{\rm Mpc}^{-1}$, a typical approach is to set
the target number of observed galaxies such that the number density satisfies
$n_gP_g(k_{\rm max})\gtrsim 1$, where $k_{\rm max}$ is the maximum wavenumber
below which we wish to measure $P_g(k)$. As a result, the uncertainty
at smaller $k$ is totally dominated by sample variance.}
At that point, the only way to reduce the uncertainty is to cover more volume.

The exact definition of $V_{\rm survey}$ requires some thoughts.
Suppose that we divide a large volume, $V_{\rm survey}$, into smaller
regions each having a volume of $V_{\rm small}$, which are separated
by distances comparable to the size of each volume. (When the smaller
non-overlapping regions are embedded in the larger volume, $V_{\rm survey}$,
the sum of $V_{\rm small}$ is smaller than $V_{\rm survey}$.) Is the survey
volume $V_{\rm survey}$, or the sum of $V_{\rm small}$? The answer
depends on the wavenumbers at which we measure $P_g(k)$. If we are
interested in measuring the power spectrum on length scales much
smaller than the size of smaller volumes, $k\gg 2\pi V_{\rm small}^{-1/3}$,
then the survey volume would be the sum of $V_{\rm small}$. On the other hand,
to reconstruct a plane-wave fluctuation having a given wavenumber,
one would not need to have a full sample of the plane wave. The Nyquist
sampling theorem states that one can completely reconstruct a plane-wave
fluctuation as long as it is sampled at the rate twice the wavenumber of
the plane wave. Therefore, if we choose the distribution of smaller
regions properly, we can reconstruct long-wavelength fluctuations
by sparsely sampling $V_{\rm survey}$ without loss of information; thus,
the survey volume in this case is equal to  $V_{\rm survey}$ despite the fact
that the observations cover only a fraction of $V_{\rm survey}$. 

A number of questions arise. For example, what is the optimal
distribution of smaller regions? Should we distribute them regularly, or
randomly? In this paper, we shall address questions related to this
``sparse sampling method'' as applied to galaxy redshift
surveys. Specifically, we shall discuss the sparse sampling method within
the context of HETDEX \cite{hill/etal:2008}; however, our discussion
is sufficiently general so that it can be applied to other galaxy
surveys measuring the power spectrum. Similar issues have been studied
in \cite{blake/etal:2006,paykari/jaffe:prep}.

This paper is organized as follows.
In section~\ref{sec:pk}, we construct a relation between the observed
galaxy power spectrum and the underlying one, making clear how the
selection  function of the sparse sampling enters into the relation.
In section~\ref{sec:1d}, we analyze a toy one-dimensional example to
understand the basic properties of the power spectrum measured from
sparsely-sampled density fields. 
In section~\ref{sec:2d}, we investigate a toy two-dimensional example to
explore the effects of shapes and orientations of observed regions.
In section~\ref{sec:3d}, we use log-normal realizations of semi-realistic
density fields in three dimensions to investigate the remaining effects
of sparse sampling in detail. We also study how sparse sampling of a
realistic galaxy survey may affect the observed Baryon Acoustic Oscillations
(BAOs) and the constraints on cosmological parameters.
In section~\ref{sec:2pcf}, we study the effect of sparse sampling
on the two-point correlation function.
We conclude in section~\ref{sec:conclusion}.
In appendix~\ref{sec:perturbation}, we derive the one-dimensional window
function effect for Gaussian perturbations to the regularly-sparse sampling.
In appendix~\ref{sec:lognormal}, we describe our log-normal simulations
of density fields.
In appendix~\ref{sec:gaussian}, we present our Gaussian realizations for
the power spectrum with window functions.

\section{Power spectrum and window function}
\label{sec:pk}
Galaxy redshift surveys measure three-dimensional locations of galaxies,
from which one can determine two-point correlation functions.
The two-point correlation function in real space can be estimated
from the observed locations directly, while computation of the two-point
correlation function in Fourier space (power spectrum) using a Fast
Fourier Transform 
requires an estimate of local number density of galaxies at regular grid points.
Using an appropriate density assignment scheme (such as the nearest-grid-point (NGP)
density assignment; the cloud-in-cell (CIC) density assignment; etc. See
\cite{{jing:2005}}), one can measure local density fields of observed
galaxies, $n_g({\bf r})$.  

The observed number density is a product of the true, underlying number
density of galaxies of a given population for a given limiting magnitude
or line flux, $n({\bf r})$, and the so-called ``selection function,'' $W({\bf r})$:
\begin{equation}
 n_g({\bf r})=n({\bf r})W({\bf r}).
\end{equation}
The observed density fluctuation is defined as (see eq.~2.1.3 of
\cite{feldman/kaiser/peacock:1994}\footnote{We are equally weighting
each of the observed galaxies by setting the weight to
be unity, i.e., $w({\bf r})\equiv 1$ in the notation of
Feldman-Kaiser-Peacock (FKP; \cite{feldman/kaiser/peacock:1994}).})
\begin{equation}
 \delta_g({\bf r})\equiv \frac{n_g({\bf r})-\bar{n}_g({\bf r})}
 {\left[\int d^3r~\bar{n}^2_g({\bf r})\right]^{1/2}}.
\label{eq:densitycontrast}
\end{equation}
Here, $\bar{n}_g({\bf r})\equiv \bar{n}(z)W({\bf r})$ is an estimate of
the {\it local} mean number density of galaxies, and $\bar{n}(z)$
is the global mean of $n({\bf r})$ at a given redshift calculated from,
e.g., a controlled field where $W({\bf r})$ is normalized to be unity. 

In this paper, we shall focus only on the spherically-averaged
(monopole) power spectrum.\footnote{The observed power spectrum can be
expanded in series of Legendre polynomials, ${\cal P}_L(\mu)$, as
$\hat{P}_g(k,\mu)=\sum_L\hat{P}_{g,L}(k){\cal P}_L(\mu)$, where $\mu$ is
the cosine between the line-of-sight and the tangential directions.
The inverse relation is $\hat P_{g,L}(k)=\frac{2L+1}{2}\int_{-1}^1 d\mu
\hat{P}_g(k,\mu){\cal P}_L(\mu)$.
The term corresponding to $L=0$ (i.e., $\hat{P}_g(k,\mu)$ averaged over
$\mu$) is called the monopole.} Fourier transforming $\delta_g$,
we calculate the observed power spectrum, $\hat{P}_g(k)$, by
\begin{equation}
 \hat{P}_g(k)\equiv \frac1{N_k}\sum_{i=1}^{N_k} \left|\delta_g({\bf k}_i)\right|^2,
\label{eq:Pg_hat}
\end{equation}
where $N_k$ is the number of Fourier meshes for which $|{\bf k}_i|$ lies
within a given wavenumber bin, i.e., $k-\delta k/2\le
|{\bf k}_i|<k+\delta k/2$.  The expectation value of $\hat{P}_g(k)$ is then
given by (see eq.~2.1.6 of \cite{feldman/kaiser/peacock:1994})
\begin{equation}
 \langle\hat{P}_g(k)\rangle=\frac{1}{W_{\rm sq}}
 \int\frac{d^3q}{(2\pi)^3}~P({\bf q})|W({\bf k}-{\bf q})|^2+P_{\rm shot},
\label{eq:Pconv_def}
\end{equation}
where $P({\bf q})$ is the true, underlying power spectrum of galaxies of
a given population for a given limiting magnitude or line flux,
$W_{\rm sq}$ is the normalization factor given by
$W_{\rm sq}\equiv\int d^3r~\bar n^2(z)W^2({\bf r})$,
and $P_{\rm shot}$ is the shot noise:
\begin{equation}
 P_{\rm shot}\equiv\frac{\int d^3r~\bar n(z)W^2({\bf r})}{\int d^3r~\bar n^2(z)W^2({\bf r})}.
\end{equation}
Here in the derivation, we have ignored the effects of density assignments.
However, in our numerical results, we do take the density assignments into
account following the procedures in ref.~\cite{jing:2005}.

In the sparse-sampling approach, there are gaps (i.e., $W({\bf r})=0$)
between observed fields. Therefore, our goal is to determine how the
sparse sampling affects the observed power spectrum and its
uncertainties. For simplicity, throughout this paper, we shall assume
that 
$\bar n(z)$ is independent of $z$, i.e. $\bar n(z)=\bar n$. Thus
$W_{\rm sq}=\bar n^2\int d^3r~W^2({\bf r})=(2\pi)^{-3}\bar n^2\int d^3k~|W({\bf k})|^2$.

\section{One-dimensional study}
\label{sec:1d}
\subsection{Regularly-spaced sparse sampling}
\begin{figure}[t]
\centering{
\includegraphics[width=1\textwidth]{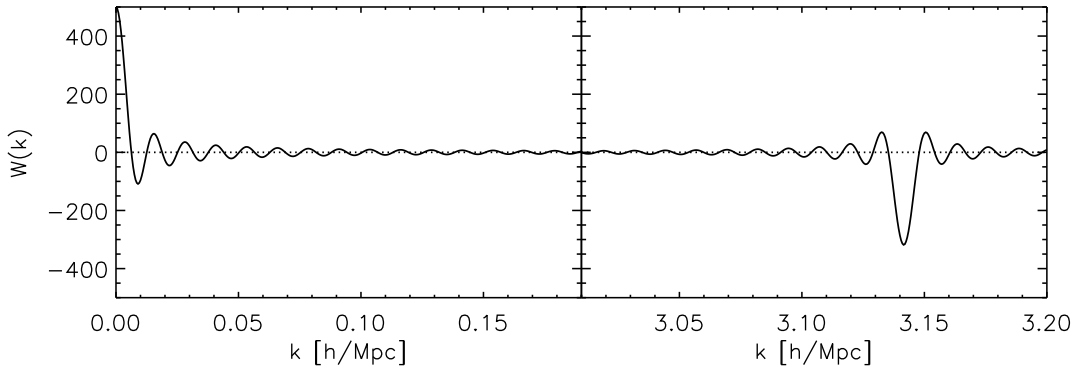}
}
\caption{(Left panel) One-dimensional window function of a
 regularly-spaced sparse sampling, $W(k)$ (see eq.~\ref{eq:Wk1d_sinc}),
 for $0\le k\le0.2~h~{\rm Mpc}^{-1}$. 
 (Right panel) Same as the left panel, but for
 $3.0\le k\le3.2~h~{\rm Mpc}^{-1}$. The survey parameters
 are $L=1000~h^{-1}~{\rm Mpc}$, $d=1~h^{-1}~{\rm Mpc}$, and $r=2~h^{-1}~{\rm Mpc}$.
}
\label{fig:Wk_1d}
\end{figure}

First, let us examine a toy example of a one-dimensional density field.
Suppose that we observe $N$ one-dimensional non-overlapping regions,
each having a size of $d$. The selection function is given by the sum
of $N$ top-hat functions: $W(x)=\sum_{m=1}^N \theta(d/2-|x-x_{c,m}|)$,
where $\theta(y)=1$ for $y\ge 0$ and 0 otherwise, and $x_{c,m}$ is the
position of the center of the $m^{\rm th}$ region. Fourier-transforming
the selection function, one finds $W(k) = d\sum_{m=1}^{N}{\rm sinc}(kd/2)e^{-ikx_{c,m}}$.
(${\rm sinc}(x)=\sin(x)/x$.) For regularly-spaced regions with a separation
between the centers given by $r$, one finds 
\begin{equation}
 W(k)=\frac{dL}{r}
 \frac{{\rm sinc}(kd/2){\rm sinc}(kL/2)}{{\rm sinc}(kr/2)},
\label{eq:Wk1d_sinc}
\end{equation}
where $L=Nr$ is the survey length. The one-dimensional window function
is shown in figure~\ref{fig:Wk_1d} for survey parameters of
$L=1000~h^{-1}~{\rm Mpc}$, $d=1~h^{-1}~{\rm Mpc}$, and $r=2~h^{-1}~{\rm Mpc}$.

For wavelengths much greater than the size of the smaller regions, $kd\ll 1$,
and the separation between the region centers, $kr\ll 1$,
the window function becomes $W(k)\propto {\rm sinc}(kL/2)$,
which is what one would expect for a contiguous survey ($r=d$).

The key behavior of this window function is that it diverges at $kr=2m\pi$
where $m$ is an arbitrary non-zero integer. This property may bias the
observed power spectrum, and the effect can be quantified by the
one-dimensional version of eq.~\ref{eq:Pconv_def}, i.e.,
\begin{equation}
 \langle\hat{P}_g(k)\rangle=\frac1{W_{\rm sq}}\int\frac{dq}{2\pi}~P(|q|)|W(k-q)|^2,
\label{eq:Pconv_1d}
\end{equation}
as $\langle\hat{P}_g(k)\rangle$
receives contributions from $P(k)$ as well as from $P(|k\pm 2m\pi/r|)$.
(We have ignored the shot noise here for simplicity.)

However, there is an easy solution to this problem. When we compute density
fields from locations of galaxies, we assign density values to meshes
(which we call ``density meshes''). If we choose the size of density meshes,
$H=L/N_{\rm mesh}$, such that the Nyquist frequency, $k_{Nyq}=\pi/H$, is
smaller than $2\pi/r$, then the above integral receives contributions
from the main peak ($k-q=0$) but does not receive contributions
from the other peaks ($k-q=\pm2\pi/r$) of the window
function, and thus 
$\langle \hat{P}_g(k)\rangle-P_{\rm shot}$ becomes unbiased, i.e.,
regularly-spaced 
sparse sampling returns an unbiased power spectrum  when the size of the
density mesh is so large/coarse ($H>r/2$) that it does not resolve
separations between 
observed regions.\footnote{The observed power spectrum does not receive the
oscillatory features in the window function because we choose the wavenumber as
$k=mk_F$, where $k_F=2\pi/L$ is the fundamental frequency and $m$ is an arbitrary
integer. Then, ${\rm sinc}(mk_FL/2)={\rm sinc}(m\pi)=0$ except for $k=2\pi/r$.} 
Mathematically, eq.~\ref{eq:Pconv_1d} is modified as
\begin{equation}
 \langle\hat{P}_g(k)\rangle=\frac1{W_{\rm sq,mesh}}
 \int\frac{dq}{2\pi}~P(|q|)|W(k-q)|^2|W_{\rm mesh}(k-q)|^2,
\label{eq:Pconv_1d_mesh}
\end{equation}
where $W_{\rm mesh}(k)$ is the Fourier transform of the density assignment
scheme and $W_{\rm sq,mesh}=\int\frac{dk}{2\pi}~|W(k)|^2|W_{\rm mesh}(k)|^2$.
For CIC, $W_{\rm mesh}(k)={\rm sinc}^2(kH/2)$. Therefore, all peaks in the
window function with wavenumber greater than $2\pi/H$ are
suppressed. To compute this expression, we use Fourier transform:
we first create a one-dimensional Fourier-space window function, $W(k)$, with
$d=1~h^{-1}~{\rm Mpc}$, $r=2~h^{-1}~{\rm Mpc}$, and $L=1000~h^{-1}~{\rm Mpc}$
(i.e., $N=500$). We then compute $|W(k)|^2|W_{\rm mesh}(k)|^2$,
Fourier transform it to real space, multiply the result by the Fourier
transform of the underlying power spectrum, and finally Fourier transform
the product back to obtain $\langle \hat{P}_g(k)\rangle$. 

\begin{figure}[t]
\centering{
\includegraphics[width=1\textwidth]{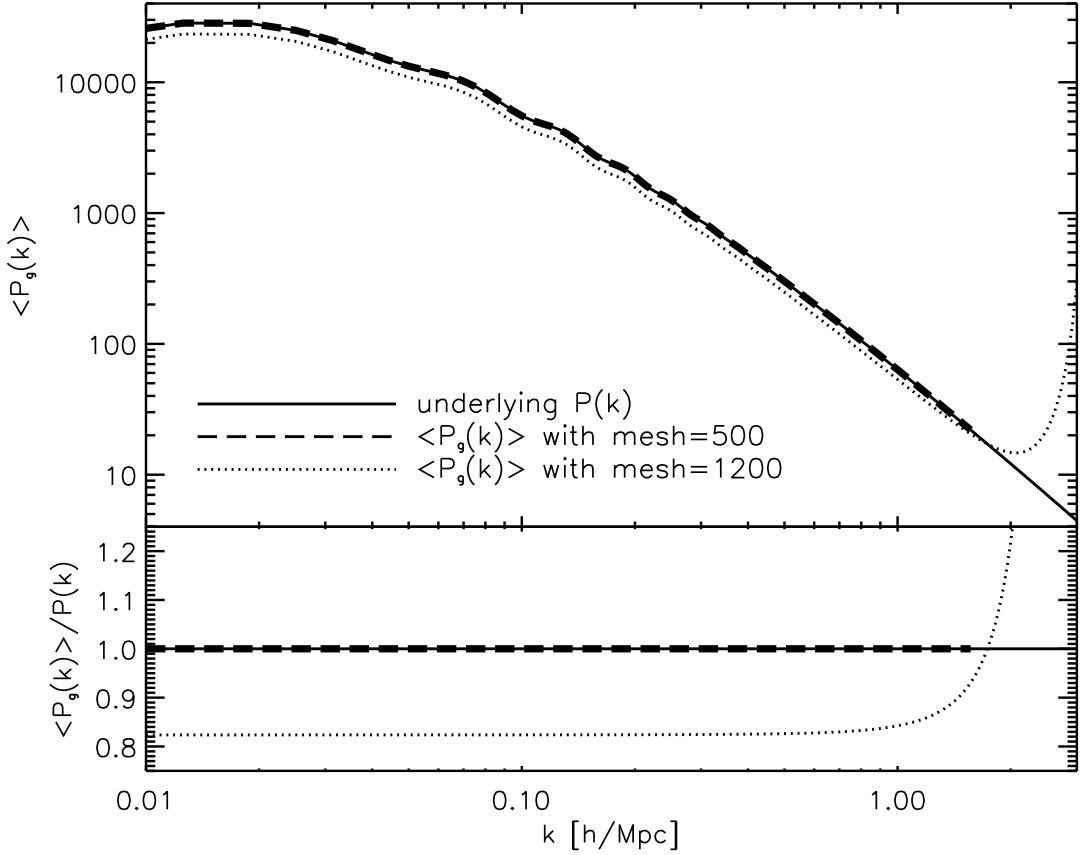}
}
\caption{(Top panel) The thick dashed and dotted lines show the expectation values of
 the observed power spectrum, $\langle \hat{P}_g(k)\rangle$, computed
 from 500 and 1200 density meshes, respectively (the total survey
 length, $L$, is divided into 500 and 1200 meshes).
 The survey parameters are $L=1000~h^{-1}~{\rm Mpc}$, $r=2~h^{-1}~{\rm Mpc}$,
 and $d=1~h^{-1}~{\rm Mpc}$.  The solid line shows the underlying power spectrum, $P(k)$.
 (Bottom panel) Ratios of $\langle \hat{P}_g(k)\rangle$ to $P(k)$.
}
\label{fig:1d_convolved_ps}
\end{figure}

In figure~\ref{fig:1d_convolved_ps}, we display two power spectra computed
from eq.~\ref{eq:Pconv_1d} with two density meshes. For 500 and 1200 meshes,
the linear sizes of each mesh are  $H=2.000$ and $0.833~h^{-1}~{\rm Mpc}$,
respectively. (Note that the survey length, $L$, is the same for both
mesh sizes.) As $r/2=1~h^{-1}~{\rm Mpc}$, the former does not resolve
separations between observed regions, while the latter does. As a result,
we find that the former yields an unbiased power spectrum, while the
latter is significantly biased.

This bias in the power spectrum we find here can be understood by
recalling that $\langle\hat{P}_g(k)\rangle$ receives contributions from
$P(k)$ as well as from $P(|k\pm 2m\pi/r|)$ with $2\pi/r\simeq 3.1~h~{\rm Mpc}^{-1}$.
As $P(k)$ is a decreasing function of $k$ for $k\gtrsim 10^{-2}~h~{\rm Mpc}^{-1}$,
contributions from $P(|k\pm 2m\pi/r|)$ are much smaller than those from
$P(k)$ for $k\ll 2\pi/r$. Therefore, eq.~\ref{eq:Pconv_1d} is approximately given by
\begin{equation}
 \langle \hat{P}_g(k)\rangle\approx \frac1{W_{\rm sq}}\frac{k_F}{2\pi}~P(k)|W(0)|^2,
\label{eq:suppression}
\end{equation}
for $k\ll 2\pi/r$. Here, $k_F=2\pi/L$ is the fundamental frequency.
For 1200 meshes, we have $W_{\rm sq}=330.533$ and $W(0)=dL/r=500$,
which gives $\langle \hat{P}_g(k)\rangle\simeq 0.824P(k)$,
in agreement with the numerical result shown in figure~\ref{fig:1d_convolved_ps}.

As we move to smaller scales, a contribution of $P(|k-2\pi/r|)$ becomes
important. For example, $P(|k-2\pi/r|)$ becomes equal to $P(k)$ for
$k=\pi/r\simeq 1.5~h~{\rm Mpc}^{-1}$, in agreement with the result in
figure~\ref{fig:1d_convolved_ps}. At $k>1.5~h~{\rm Mpc}^{-1}$, the contribution of
$P(|k-2\pi/r|)$ exceeds that of $P(k)$, enhancing $\langle P_g(k)\rangle$ above $P(k)$.

\subsection{Perturbations to regularly-spaced sparse sampling}
Due to various observational constraints, it is often not possible
to have perfectly regular separations between observed regions.
To study the effect of deviations from regular spacing, we perturb the
center of the $a^{\rm th}$ observed region by $\epsilon_a$, 
where $\epsilon_a$ is a Gaussian variable with zero mean,
$\langle\epsilon_a\rangle=0$, and the correlation function given by
$\langle\epsilon_a\epsilon_b\rangle=\sigma_{\epsilon}^2\delta_{ab}$. 
The expectation value of the window function squared is given by
\begin{eqnarray}
 \langle\left|W(k)\right|^2\rangle&=&
 \left[d\ {\rm sinc}\left(\frac{kd}{2}\right)\right]^2
 \sum_{a=1}^N\sum_{b=1}^Ne^{-ik\bar{x}_a}e^{ik\bar{x}_b}
 \langle e^{-ik\epsilon_a}e^{ik\epsilon_b}\rangle \nonumber\\
 &=&\left[d\ {\rm sinc}\left(\frac{kd}{2}\right)\right]^2\left[N+
 e^{-k^2\sigma_{\epsilon}^2}\sum_{a\neq b}e^{-ik(\bar{x}_a-\bar{x}_b)}\right],
\label{eq:Wk_gaussian_pert}
\end{eqnarray}
where $\bar{x}_a$ denotes the unperturbed, regularly-spaced position of
the $a^{\rm th}$ region. The derivation of this result is given in
appendix \ref{sec:perturbation}.

As the perturbations of the center of the observed regions
would suppress the observed power spectrum on large scales,
we quantify the effect of non-regularity by computing the large-scale
suppression of the power spectrum as a function of the standard
deviation of perturbations, $\sigma_\epsilon$. We compute $W_{\rm sq}$
by the discrete sum as
$W_{\rm sq}=\frac{k_F}{2\pi}\sum_{n=-n_{Nyq}}^{n_{Nyq}}\langle|W(nk_F)|^2\rangle
|W_{\rm mesh}(nk_F)|^2$,
where $n_{Nyq}$ is equal to the Nyquist frequency divided by the
fundamental frequency, $k_F=2\pi/L$.

\begin{figure}[t]
\centering{
\includegraphics[width=1\textwidth]{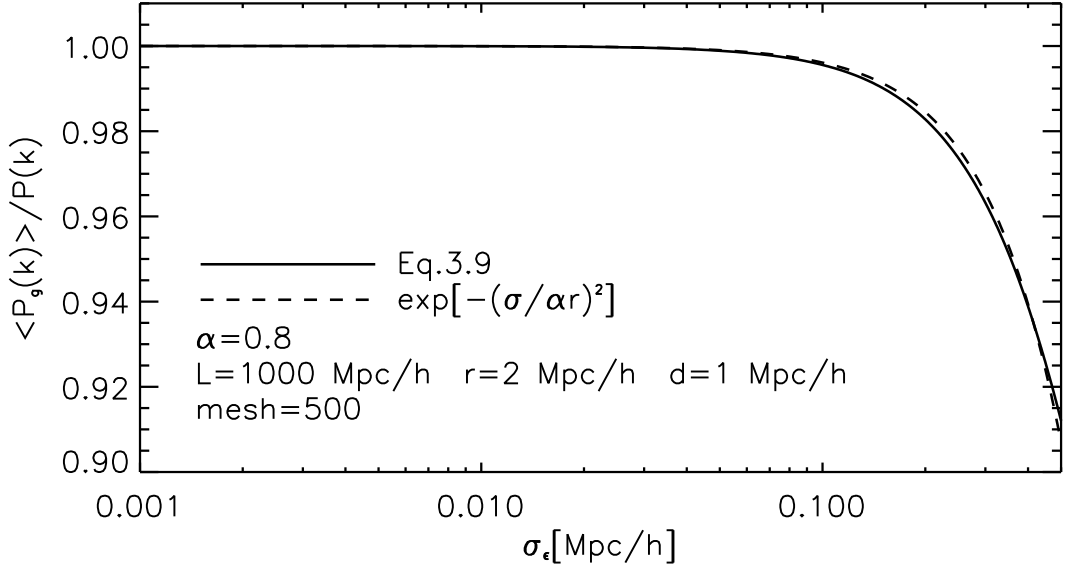}
}
\caption{
 Ratio of the expectation value of the observed power spectrum,
 $\langle\hat{P}_g(k)\rangle$, to the underlying power spectrum, $P(k)$,
 computed from eq.~\ref{eq:fsupp_gaussian_pert} for $k\ll 2\pi/r$ as a
 function of the magnitude of perturbations to the positions of the centers
 of observed regions, $\sigma_{\epsilon}$. The survey parameters are
 $L=1000~h^{-1}~{\rm Mpc}$, $r=2~h^{-1}~{\rm Mpc}$, and $d=1~h^{-1}~{\rm Mpc}$,
 and 500 density meshes are used.
 The solid line is computed from eq.~\ref{eq:fsupp_gaussian_pert}.
 while the dashed line is an approximate fitting formula,
 $\exp\left[-(\sigma_{\epsilon}/\alpha r)^2\right]$, where $\alpha=0.8$.
}
\label{fig:1d_large_scale_sup}
\end{figure}

For $k=0$ ($n=0$), the last term of eq.~\ref{eq:Wk_gaussian_pert} becomes
\begin{equation}
 \sum_{a\neq b}e^{-ik(\bar{x}_a-\bar{x}_b)}=\sum_{a\neq b}1
 =2\sum_{a=1}^{N-1}\sum_{b=0}^{a-1}1=N^2-N.
\label{eq:sum_exp_keq0}
\end{equation}
For $k\neq 0$ ($n\neq 0$), 
\begin{eqnarray}
 \sum_{a\neq b}e^{-ik(\bar{x}_a-\bar{x}_b)}&=&\sum_{a\neq b}e^{-ikr(a-b)}
 =2\sum_{a=1}^{N-1}\sum_{b=0}^{a-1}\cos\left[kr(a-b)\right]\nonumber\\
 &=&\frac{-1+N-N\cos(kr)+\cos(Nkr)}{-1+\cos(kr)},
\label{eq:sum_exp_kneq0}
\end{eqnarray}
where we have defined $\bar{x}_a=x_0+ar$ ($0\le a\le N-1$) and $x_0=(-L+r)/2$.
Recalling $L=Nr$, one finds $\cos(Nkr)=\cos(nk_FL)=\cos(2n\pi)=1$; thus, the
right hand side of eq.~\ref{eq:sum_exp_kneq0} is equal to $-N$. Using these
results, 
\begin{eqnarray}
 W_{\rm sq}=\frac{k_F}{2\pi}\Bigg\lbrace d^2N^2+2d^2N
 \sum_{n=1}^{n_{Nyq}}{\rm sinc}^2\left(\frac{nk_Fd}{2}\right)
 \left[1-e^{-(nk_F\sigma_{\epsilon})^2}\right]
 |W_{\rm mesh}(nk_F)|^2\Bigg\rbrace.
\label{eq:Wsq_gaussian_pert}
\end{eqnarray}
Inserting eq.~\ref{eq:Wsq_gaussian_pert} into eq.~\ref{eq:suppression},
\begin{eqnarray}
 \frac{\langle \hat{P}_g(k)\rangle}{P(k)} =\frac{N}{N+2\sum_{n=1}^{n_{Nyq}}
 {\rm sinc}^2(nk_Fd/2)\left[1-e^{-(nk_F\sigma_{\epsilon})^2}\right]
 |W_{\rm mesh}(nk_F)|^2},
\label{eq:fsupp_gaussian_pert}
\end{eqnarray}
for $k\ll 2\pi/r$. Figure~\ref{fig:1d_large_scale_sup} shows
$\langle \hat{P}_g(k)\rangle/P(k)$ as a function of $\sigma_{\epsilon}$.
The solid line is computed from eq.~\ref{eq:fsupp_gaussian_pert},
while the dashed line is an approximate fitting formula,
$\exp\left[-(\sigma_{\epsilon}/\alpha r)^2\right]$, where $\alpha=0.8$.

We find that, if perturbations are much smaller than the separation between
observed regions (i.e., $\sigma_\epsilon\ll r$), they have a negligible impact
on the power spectrum. As $\sigma_{\epsilon}$ becomes larger, the observed
large-scale power is suppressed with respect to the underlying one.
This exercise shows that the optimal strategy for sparse sampling is to
make the distribution of observed regions as regularly separated as
possible. 

\section{Two-dimensional study}
\label{sec:2d}
\subsection{Tiling the sky with shots}

We now extend the toy model to two dimensions. Suppose that we have
the focal plane of a telescope which we wish to fill with many IFUs.
Each IFU consists of densely packed optical fibers. (For HETDEX, each
IFU consists of 448 fibers.) While it would be
ideal to fill the focal plane entirely with IFUs without any gaps,
limited resources and technical practicality usually prevent doing
so. (For HETDEX, 75 IFUs are placed on the focal plane filling only a
quarter of the focal plane area.) Therefore, motivated by the study in 
the previous section, we place our IFUs such that the focal plane
contains IFUs which are separated by regular spacings.

Let us now define the term ``shot,'' as the projection
of the focal plane onto the sky. We conduct a galaxy survey
by tiling the sky with many shots. The question we wish to answer in
this section is, ``{\it how should we tile the sky with shots?}'' 

\begin{figure}[t]
\centering{
\includegraphics[width=0.32\textwidth]{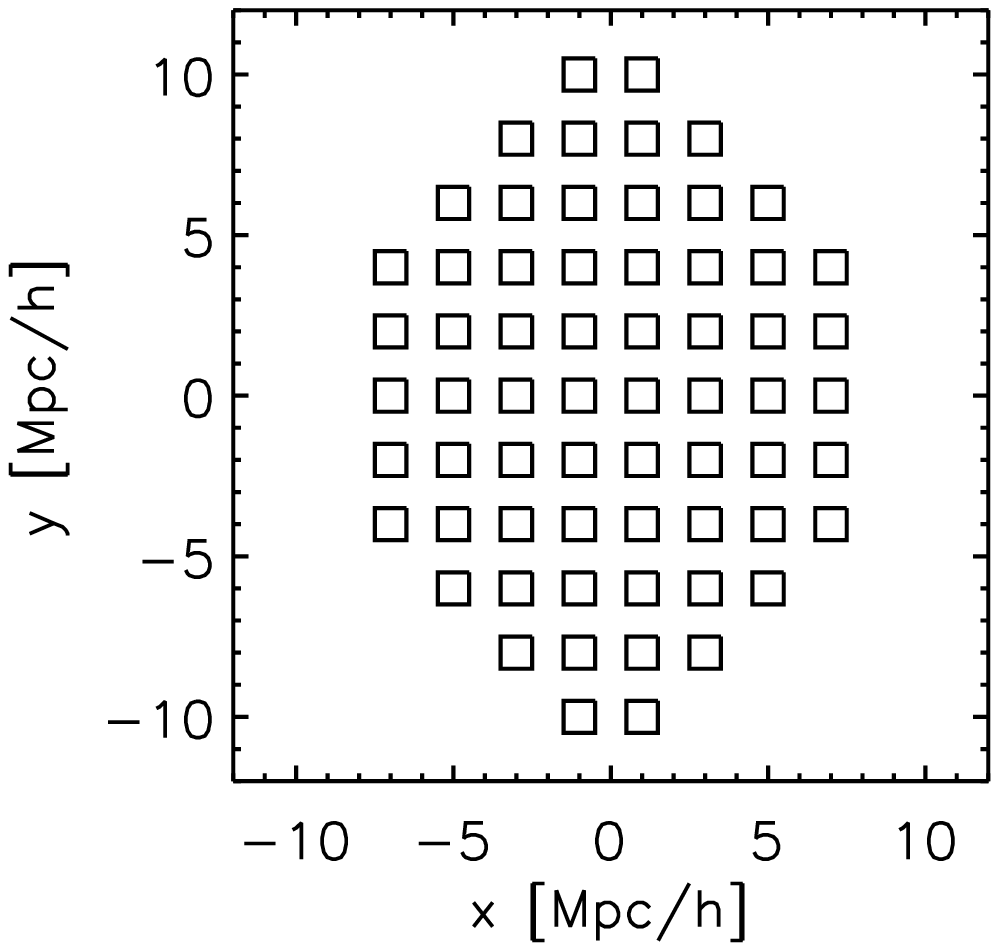}
\includegraphics[width=0.32\textwidth]{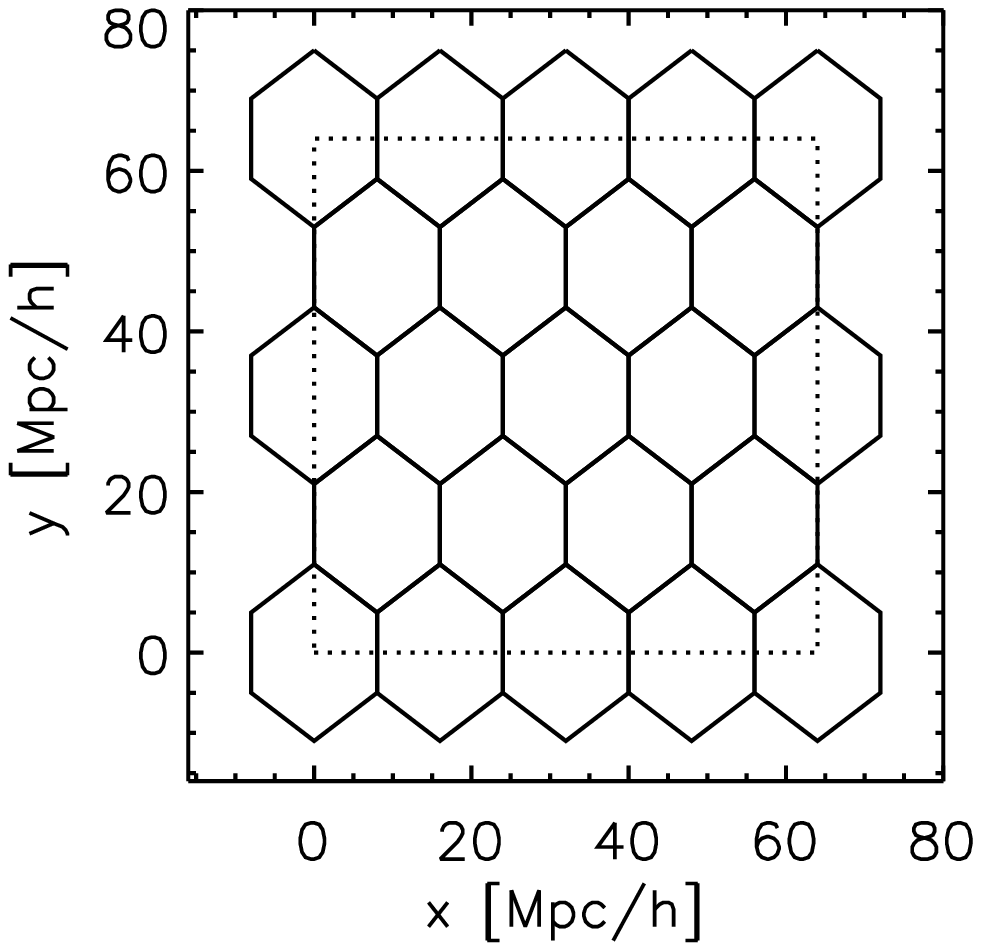}
\includegraphics[width=0.32\textwidth]{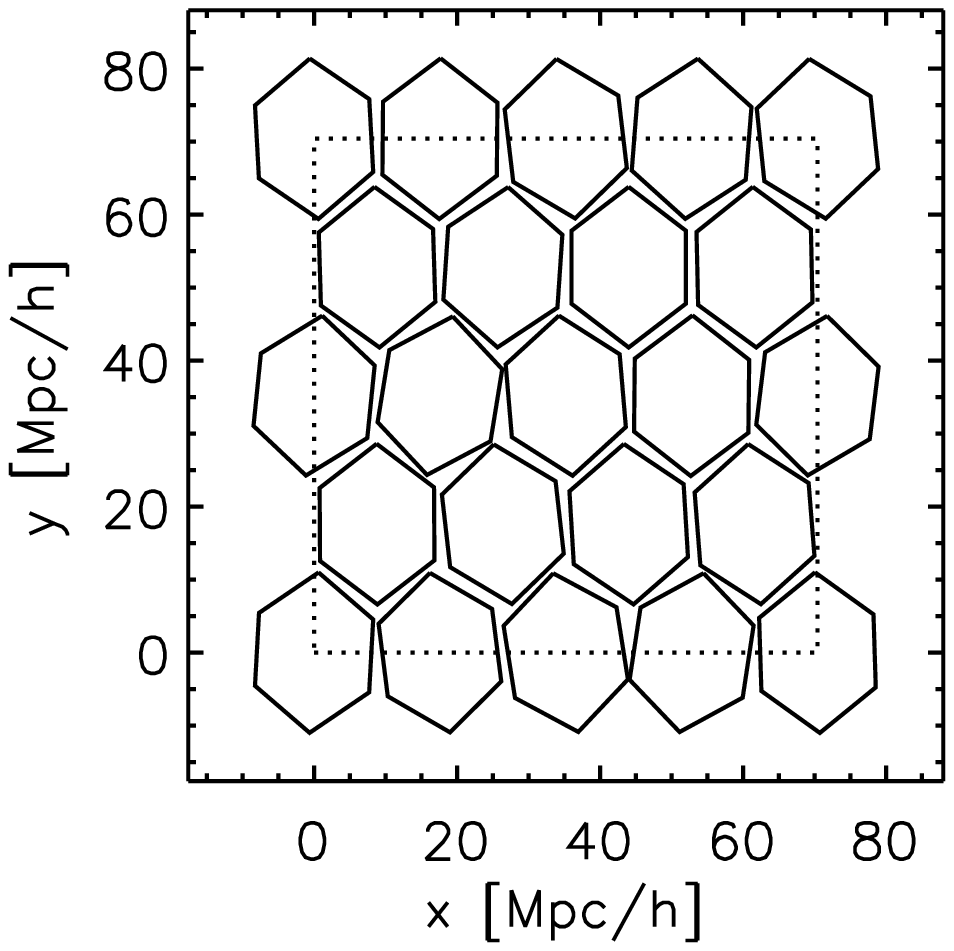}
}
\caption{(Left panel) Distribution of IFUs (shown by squares)
 for a single ``shot,'' which is the focal plane projected onto
 the sky. There are 64 IFUs in this example. Each IFU covers
 the comoving size of $1~h^{-1}~{\rm Mpc}$, and they are regularly
 separated by $2~h^{-1}~{\rm Mpc}$. (Middle panel) Survey area
 tiled without any gaps between shots (each shot consists of 64 IFUs as
 shown in the left panel). The dotted line shows the survey
 boundary, whose area is $64~h^{-1}~{\rm Mpc}\times 64~h^{-1}~{\rm Mpc}$.
 The survey area shown here is 10 times smaller than what is
 actually used by the calculations presented in the main text.
 Separations between the shot centers are $16~h^{-1}~{\rm Mpc}$ along $x$
 and $y$ directions. (Right panel) Survey area tiled
 {\it with} gaps and random rotations. Separations between the
 shot centers are $17.6~h^{-1}~{\rm Mpc}$. Orientations of shots
 are rotated randomly by angles between $-10^{\circ}$ and $+10^{\circ}$.
 The survey area is $70.4~h^{-1}~{\rm Mpc}\times 70.4~h^{-1}~{\rm Mpc}$.
}
\label{fig:64ifus}
\end{figure}

The left panel of figure~\ref{fig:64ifus} shows one example.
Each small square represents one IFU. Objects will only be
detected if they fall within one IFU. The IFUs are approximately
arranged in a hexagonal pattern. There are (at least) two reasons
to prefer a hexagon over a square or a circle. First, due to the
HET specifics, the image quality and throughput tend to degrade
toward the edges of the focal plane. Therefore, a circle or a
hexagon is preferred over a square, as the length of diagonal
of a square is longer than that of the side of a square. Second,
we wish to perform a galaxy survey by tiling the sky. As one cannot
completely tile the sky by circles without significant overlaps, a
square or a hexagon is preferred over a circle. These constraints
identify a hexagon as the best choice.

In the plane-parallel approximation, we set the
center-to-center separation between IFUs to be $2~h^{-1}~{\rm Mpc}$,
and the size of each IFU to be $d=1~h^{-1}~{\rm Mpc}$. ($1~h^{-1}~{\rm Mpc}$
corresponds to roughly 0.88 arcminute at $z=2.2$.)  With this
configuration, the filling fraction (ratio of the area of IFUs to
the area of a shot) is roughly 25\%, so approximately $1/4$ of
galaxies within a single shot will be observed. 

For a single IFU, the real-space selection function is a two-dimensional
top-hat function. In Fourier space, the window function of a single IFU becomes
\begin{equation}
 W_{\rm IFU}({\bf k})=d^2{\rm sinc}\left(\frac{k_xd}{2}\right){\rm sinc}\left(\frac{k_yd}{2}\right) \ .
\end{equation}
The window function of a single shot (see the left panel of
figure~\ref{fig:64ifus}) is then given by
$W_{\rm shot}({\bf k})=\sum_{m=1}^{64} W_{\rm IFU}({\bf k})e^{-i{\bf k}\cdot{\bf r}_m}$,
where ${\bf r}_m$ is the position of the $m^{\rm th}$ IFU with
respect to the center of the shot. 

We tile the survey area by many shots.
Suppose that we use $N$ shots to fill a given survey area,
and the $n^{\rm th}$ shot is located at ${\bf L}_n$
with respect to the origin of the survey area.
Ignoring the possibility that each shot can 
be rotated (we shall return to this possibility shortly),
the final two-dimensional window function is 
\begin{equation}
 W({\bf k})=\sum_{n=1}^N W_{\rm shot}({\bf k})e^{-i{\bf k}\cdot{\bf L}_n}
 =W_{\rm IFU}({\bf k})\sum_{n=1}^N\sum_{m=1}^{64}e^{-i{\bf k}\cdot({\bf L}_n+{\bf r}_m)} \ ,
\label{eq:2d_wk_norot}
\end{equation}
where ${\bf L}_n+{\bf r}_m$ is the position of the $m^{\rm th}$ IFU in
the $n^{\rm th}$ shot. Once the window function is computed,
one can calculate the observed power spectrum by
the two-dimensional version of eq.~\ref{eq:Pconv_1d_mesh}, i.e.,
\begin{equation}
 \langle \hat{P}_g(k)\rangle=\frac1{W_{\rm sq}}
 \int\frac{d^2q}{(2\pi)^2}~P(|{\bf q}|)|W({\bf k}-{\bf q})|^2|W_{\rm
 mesh}({\bf k}-{\bf q})|^2 \ ,
\label{eq:Pconv_2d}
\end{equation}
where $W_{\rm mesh}({\bf k})={\rm sinc}^2(k_xH/2){\rm sinc}^2(k_yH/2)$ for CIC.
We use this equation to study how the
observed power spectrum is biased for a given distribution of shots in the sky.

First, let us consider the case in which shots completely
cover the survey area without gaps between shots
(but there are still small separations between IFUs),
as shown in the middle panel of figure~\ref{fig:64ifus}.
For this example, the separation between shots is
$16~h^{-1}~{\rm Mpc}$ along the $x$ and $y$ directions. 
We tile the survey area by 41 shots in the odd rows and 40 shots in the
even rows, giving $41\times21+40\times20=1661$ shots in total.
The survey area is 640 ($=16\times40$) $h^{-1}$~Mpc by 640~$h^{-1}$~Mpc.
For the further analysis we only consider a square-shaped subregion
(shown by the dotted lines in the middle panel of
figure~\ref{fig:64ifus}) of
the actually covered region. This avoids the complications of the
unevenly covered regions around the survey edges and simplifies the
window function. We count the IFUs if their centers are within the
subregion. If IFUs are partially covered at the edge of the subregion,
the window functions of the IFUs would vary, but this effect is
negligible. 

To compute the window function, we apply eq.~\ref{eq:2d_wk_norot}. We
then convolve 
the underlying power spectrum with the window function using the Fourier
transformation: we multiply the Fourier transform of
$|W({\mathbf{k}})|^2$ by the 
Fourier transform of the underlying power spectrum, and
compute the inverse Fourier transform of the product. 
We set the size of the density
mesh to be $2~h^{-1}~{\rm Mpc}$, which is large enough so that
the separation between IFUs within a shot does not bias the
observed power spectrum according to our one-dimensional study
given in the previous section.

\begin{figure}[t]
\centering{
\includegraphics[width=1\textwidth]{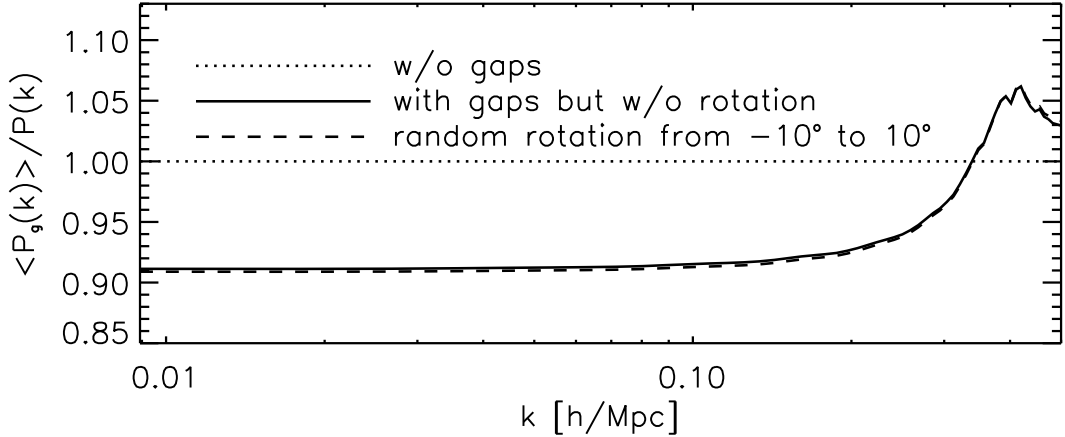}
}
\caption{Ratios of the observed power spectrum, $\langle\hat{P}_g(k)\rangle$,
 to the underlying power spectrum, $P(k)$. The dotted,
 solid, and dashed lines show  
 $\langle\hat{P}_g(k)\rangle/P(k)$ with tiled shots without gaps, shots with
 gaps but no rotation, and random rotations between $-10^{\circ}$ and
 $+10^{\circ}$ (i.e., the right
 panel of figure~\ref{fig:64ifus}), respectively.}
\label{fig:2d_convolved_pk}
\end{figure}

Since there are no gaps between shots, the survey area is sampled
regularly by IFUs which are separated by $2~h^{-1}~{\rm Mpc}$. Therefore,
as expected from our one-dimensional study, the observed power
spectrum is unbiased, which is shown as the dotted line in
figure~\ref{fig:2d_convolved_pk}. However, while the observed
power spectrum is unbiased by sparse sampling, the uncertainties
of the power spectrum increase, as the number of observed galaxies
is less than what would be obtained from a contiguous survey.
We shall investigate the effect of sparse sampling on the uncertainties
of the power spectrum in section \ref{sec:3d}.

Now let us consider the case in which there are small gaps between
shots. As an example, we increase the separation between shots to
$|\delta {\mathbf r}|=17.6~h^{-1}~{\rm Mpc}$ along the $x$ and $y$ directions,
which is 10\% larger than the previous case in which the shots completely fill
the survey area without gaps. Here, $\delta {\mathbf r}\equiv {\mathbf r}_i-{\mathbf r}_j$.
The number of shots (1661) is the same as before, and  the survey area
increases to 704 ($=17.6\times40$) $h^{-1}~{\rm Mpc}$ by $704~h^{-1}~{\rm Mpc}$.

The solid line in figure~\ref{fig:2d_convolved_pk} displays the result
of this example. We find $\sim 9$\% suppression of the power spectrum
on large scales, and an enhancement of power on small scales. The reason
for the suppression and enhancement are exactly the same as that for the
one-dimensional study: gaps between shots introduce an additional regular
spacing scale, yielding the suppression and enhancement of the observed
power spectrum with respect to the underlying one. Therefore, one should
avoid having any gaps between shots in order to minimize the window
function effect. One can estimate the magnitude of the large-scale
suppression via the two-dimensional version of eq.~\ref{eq:suppression}, i.e.,
\begin{equation}
 \langle \hat{P}_g(k)\rangle\approx \frac1{W_{\rm sq}}
 \left(\frac{k_F}{2\pi}\right)^2~P(k)|W(0)|^2 \ .
\label{eq:2d_suppression}
\end{equation}
Here, $W(0)=d^2N_{\rm IFU}$ and $N_{\rm IFU}$ is the total number of
regions covered by IFUs in
the survey area. Using $N_{\rm IFU}=102720$, $W_{\rm sq}=23388$, and 
$(k_F/2\pi)=1/(704~h^{-1}$~Mpc), the above formula produces a large-scale
suppression of 0.910, which is in an excellent agreement with the result
shown in figure~\ref{fig:2d_convolved_pk}, which was obtained directly
from eq.~\ref{eq:Pconv_2d}.

\subsection{Randomly rotated shots}
Let us now consider the case in which there are small gaps between shots,
and the orientations of shots are randomly rotated, as shown in the right
panel of figure~\ref{fig:64ifus}. For a single rotated shot, the window
function is $W_{\rm shot,rot}({\bf k})=W_{\rm shot}(R[\theta]{\bf k})$,
where $R[{\theta}]$ is a rotation matrix and $\theta$ is a rotation angle
with respect to the positive $x$ direction. For $N$ shots with different
rotations, we can extend eq.~\ref{eq:2d_wk_norot} as
\begin{eqnarray}
 W({\bf k})&=&\sum_{n=1}^N W_{\rm shot,rot}({\bf k})e^{-i{\bf k}\cdot{\bf L}_n}
 =\sum_{n=1}^N W_{\rm shot}(R[\theta_n]{\bf k})e^{-i{\bf k}\cdot{\bf L}_n} \nonumber\\
 &=&\sum_{n=1}^NW_{\rm IFU}(R[\theta_n]{\bf k})e^{-i{\bf k}\cdot{\bf L}_n}
 \sum_{m=1}^{64}e^{-i(R[\theta_n]{\bf k})\cdot{\bf r}_m} \ ,
\label{eq:2d_wk_rot}
\end{eqnarray}
where $\theta_n$ is the rotation angle of the $n^{\rm th}$ shot.

The dashed line in figure~\ref{fig:2d_convolved_pk} shows the result
of the rotated shots. Here, we have rotated shots randomly by angles
between $-10^\circ$ and $+10^\circ$.\footnote{eq.~\ref{eq:2d_wk_rot}
cannot account for the overlaps between IFUs, as the overlapping area
between IFUs doubles $W({\bf r})$. If there are overlaps, one should
compute the window function numerically by generating many random
particles in the observed regions within a given survey area. When
the separations are $17.6 ~h^{-1}~{\rm Mpc}$, the hexagonal shots
do not overlap with each other as long as rotation angles lie between
$-10^{\circ}$ and $+10^{\circ}$. For HETDEX in the northern region,
we have two fixed rotation  angles, $-60^\circ$ and $+60^\circ$,
about which there are perturbations of order $10^\circ$.} We find that
the rotation affects the power spectrum only slightly ($\lesssim1$\%)
and thus is sub-dominant compared to the effect of gaps between shots.
Rotations sometimes move IFUs out of the bounded area, causing a small
change in the window function. The extra suppression on large scales
for the random rotation is due to the fewer IFUs in the survey area.
This large-scale suppression can be estimated by
eq.~\ref{eq:2d_suppression} as well.

Our two-dimensional study demonstrates that gaps between shots have
much larger impact on the observed power spectrum than small rotations.
As gaps between shots are typically larger than the size of the density mesh,
the optimal strategy for sparse sampling is to avoid having gaps between shots.

\subsection{Shot with a hole in the middle}
\begin{figure}[t]
\centering{
\includegraphics[width=0.32\textwidth]{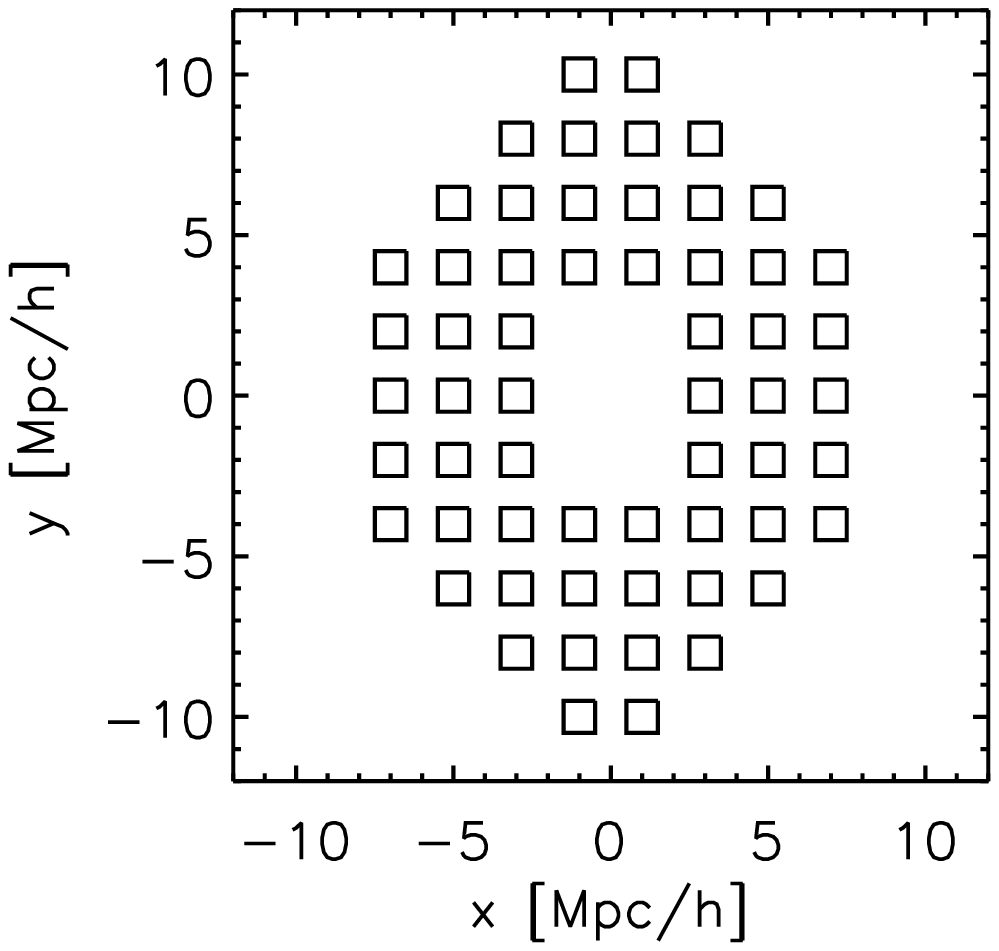}
\includegraphics[width=0.67\textwidth]{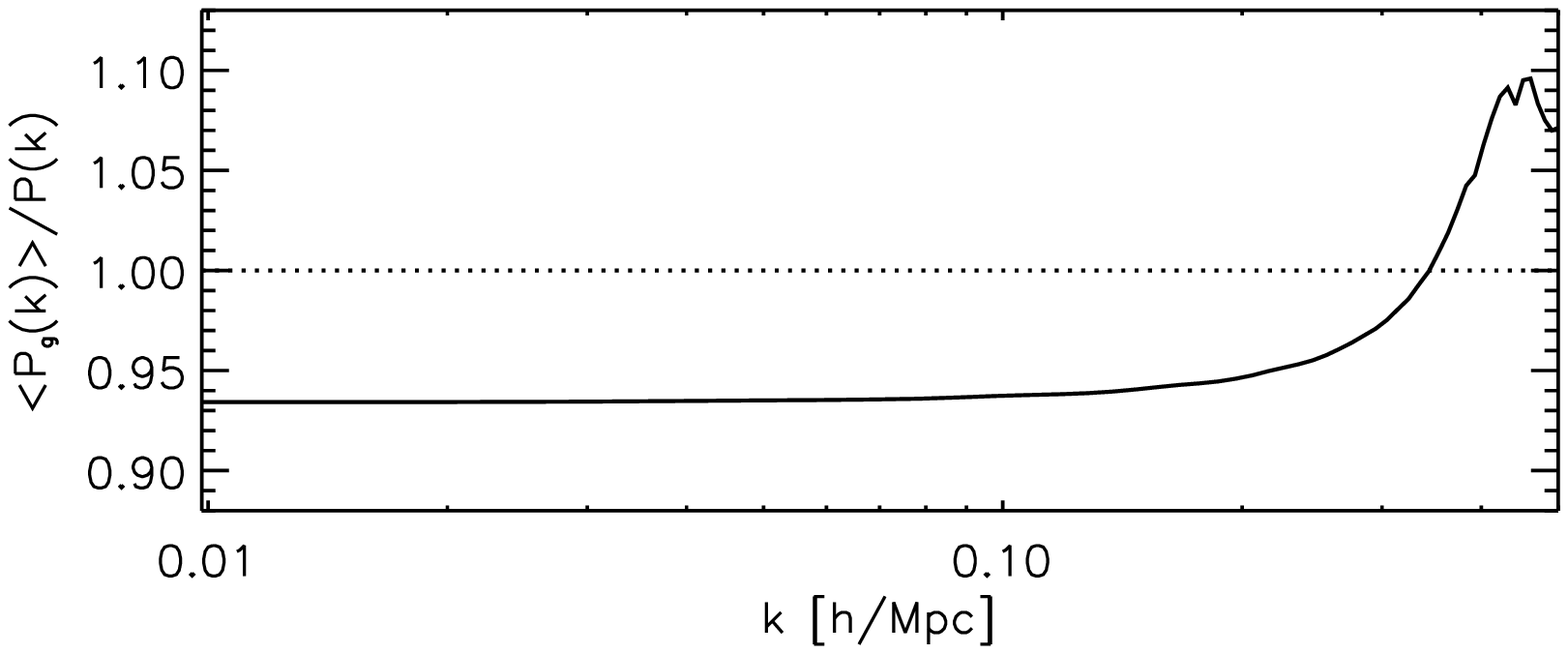}
}
\caption{(Left panel) A distribution of IFUs within a shot with a hole
 in the middle of the pattern. There are 58 IFUs. (Right panel) Ratio
 of the observed power spectrum, $\langle\hat{P}_g(k)\rangle$, to the
 underlying power spectrum, $P(k)$, for the distribution of IFUs shown
 on the left panel. There are no gaps between shots, and thus the window
 function effect is solely due to the hole in the middle. 
}
\label{fig:58ifus}
\end{figure}

Finally, let us consider a specific case for HETDEX. There will be an
instrument with higher dispersion in the center of the HET focal plane,
and the layout of IFUs would contain a hole with size of 6 IFUs in the
middle of the pattern.

The left panel of figure~\ref{fig:58ifus} shows the distribution of IFUs 
on the focal plane with no coverage in the middle of the field, and the
right panel shows the effect of the hole on the observed power spectrum.
There are no gaps between shots, but the existence of the hole creates
an additional artificial scale in the window function. For the case
explored here, in which the area of a hole is about 10\% of the focal
plane area, we find $\sim 7\%$ suppression of the power spectrum on large
scales, and an enhancement of power on small scales. The impact of the
hole in the middle of a shot is as large as that of 10\% gaps between shots. 

Again, one can estimate the magnitude of the large-scale suppression by
eq.~\ref{eq:2d_suppression}. Using $N_{\rm IFU}=93040$,
$W_{\rm sq}=22640.1$, and $k_F/(2\pi)=1/640~h^{-1}~{\rm Mpc}$,
we find the large-scale suppression of $0.933$, which is in good agreement
with the result shown in figure~\ref{fig:58ifus}. 

\section{Three-dimensional study}
\label{sec:3d}
\subsection{Effect of curvature of the sky}
We now study a semi-realistic, three-dimensional model. Suppose that
we undertake a galaxy survey whose survey footprint is bounded by
right ascension (RA), declination (DEC), and redshift. Throughout
this paper, we shall use the term ``footprint'' to denote the outermost
boundary of the survey, which contains many sub-volumes of observed regions.

As the celestial sphere is spherical, the survey volume is no longer
a cube in Cartesian coordinates. This means that one should, ideally, use the
spherical Fourier-Bessel decomposition (rather than the usual Fourier transform)
to treat density fields in radial and tangential directions separately
\cite{heavens/taylor:1994,rassat/refregier:2011,leistedt/etal:2012}.

However, in this paper we shall continue to use the fast Fourier transform
approach. The spherical Fourier-Bessel decomposition is still new in the
large-scale structure community, and much work is left to do before we
fully understand the optimal implementation of the method. 
On the other hand, the Fourier transform is easier to implement, computationally
less expensive, and is also
the conventional way to compute the power spectrum. (The Fourier
transform has been used by all of the major galaxy surveys such as
2dFGRS, SDSS, WiggleZ, and VIPERS.) 

We use a cuboidal box which
is just large enough to contain the entire survey footprint. This choice
produces a non-trivial selection function $W({\bf r})$: namely, $W({\bf
r})=1$ if ${\bf r}$ 
lies within the survey footprint, and 0 otherwise.

To study the effect of the curvature of the celestial sphere, we generate
1000 sets of mock catalogues of galaxies for a HETDEX-like survey. The survey
footprint covers $34.1^{\circ}\times7.5^{\circ}$ on the sky and a redshift
range of $1.9\le z\le 2.5$.\footnote{While the redshift range of the HETDEX
survey is $1.9\le z\le 3.5$, we choose to work with only the lower redshift
portion of the survey. The number density of Lyman-$\alpha$ galaxies
detected by HETDEX is expected to be approximately constant over the lower
redshift bin, $1.9\le z<2.5$. This bin is also more important for detecting
the effect of dark energy on the expansion rate if dark energy is a cosmological
constant. The HETDEX survey also covers an additional equatorial region in the
sky, which will not be considered in this paper, as the survey
strategies required for the northern and equatorial regions are
different. We only explore the survey strategy for the northern region
in this paper as an example (and the strategy for the equatorial region
is more straightforward).} The central coordinates of the
survey footprint in the sky are $(\mbox{RA},\mbox{DEC})=(13^{\rm h},53^{\circ})$
(J2000).

The input galaxy power spectrum for simulations is a non-linear power spectrum
at $z=2.2$ based on the third-order perturbation theory with non-linear bias
\cite{jeong/komatsu:2006,jeong/komatsu:2009}. The bias parameters are $b_1=2.2$,
$b_2=0.671$, and $P_0=72.13~h^{-3}~{\rm Mpc}^3$, and the number density of galaxies
is $\bar{n}=2.95\times 10^{-3}~h^{3}~{\rm Mpc^{-3}}$. We use log-normal realizations
of the input power spectrum to generate mock catalogues, and describe our method to
generate log-normal realizations in appendix~\ref{sec:lognormal}.

\begin{figure}[t]
\centering{
\includegraphics[width=0.48\textwidth]{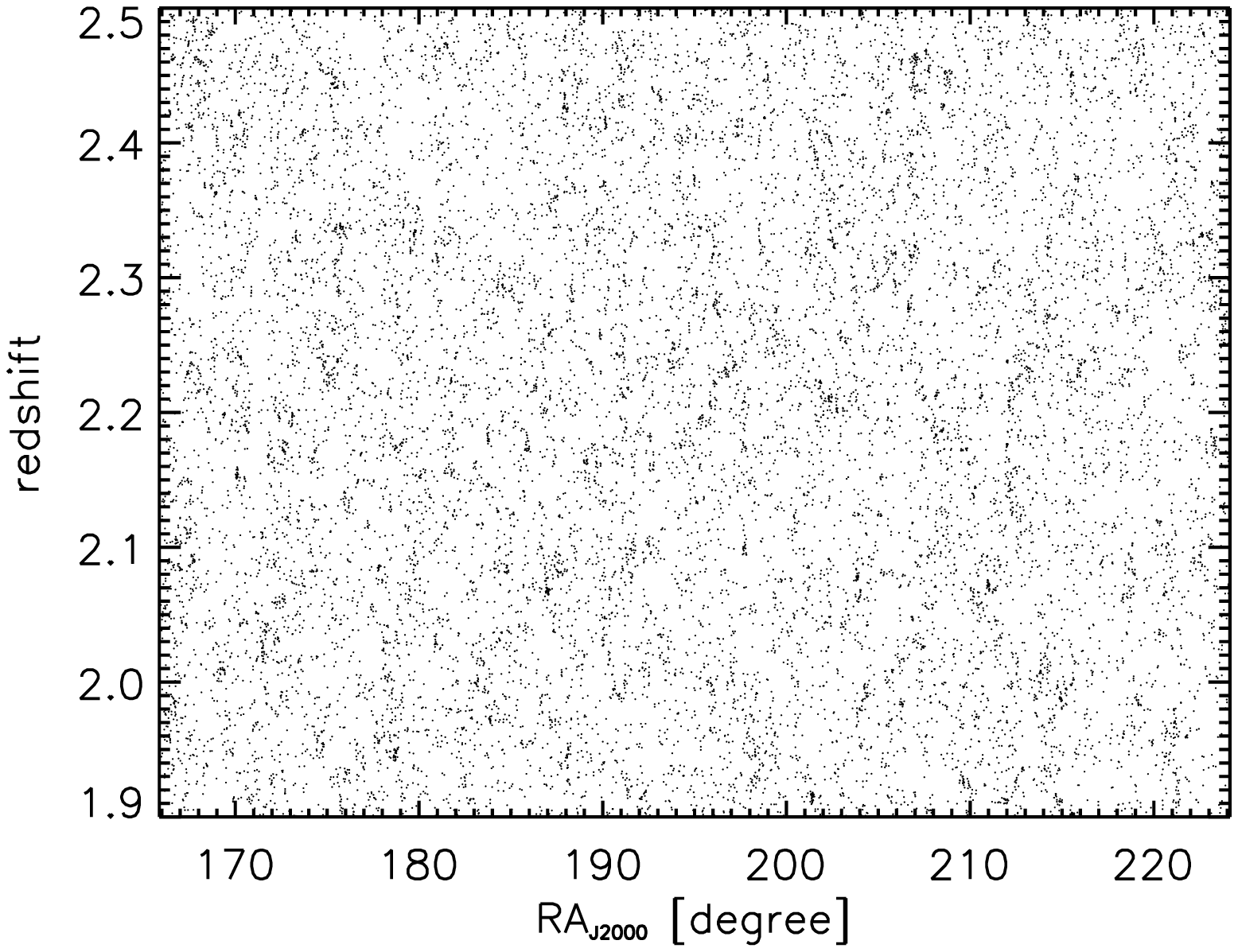}
\includegraphics[width=0.48\textwidth]{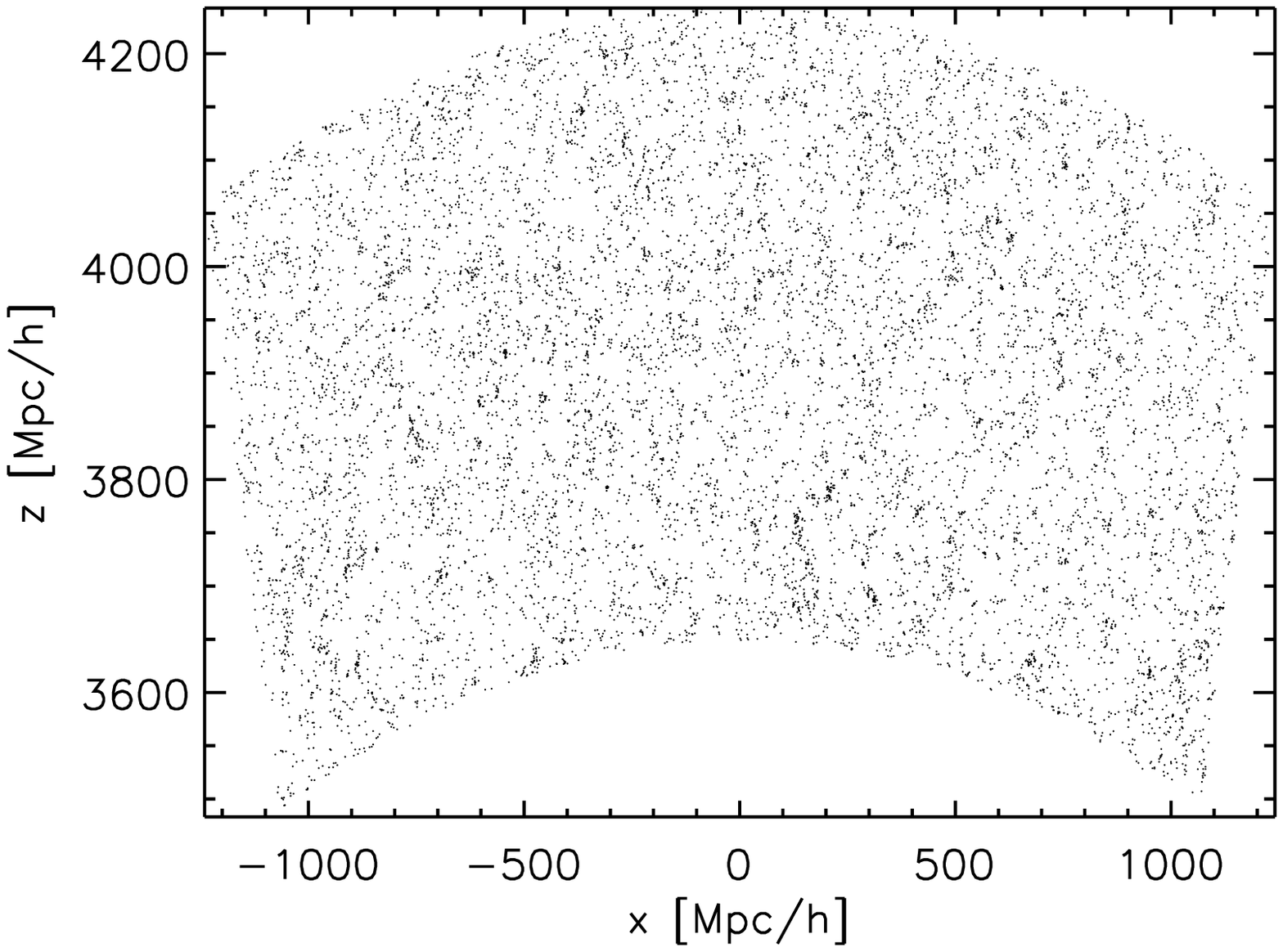}
}
\caption{
 A log-normal realization of galaxies. (Left panel) A slice on the
 RA-redshift plane with $52.974^{\circ}\le\mbox{DEC}\le 53.026^{\circ}$.
 (Right panel) A slice on the $x$-$z$ plane of the simulation box in
 Cartesian coordinates with $-1.350\le y\le1.350~h^{-1}~{\rm Mpc}$.
}
\label{fig:lognormal_galaxies}
\end{figure}

Once galaxies are created in the cuboidal simulation box in Cartesian coordinates, 
we select {\it all} galaxies lying within the survey footprint.\footnote{For
this subsection, we are not yet doing a sparse sampling but are performing
a contiguous survey with completely filled focal plane and shots without
any gaps. The sparse sampling of three-dimensional density fields will be
explored in the next subsection.} The side lengths of the simulation box
along $x$, $y$, and $z$ directions are $L_x=2481.94~h^{-1}~{\rm Mpc}$,
$L_y=540.122~h^{-1}~{\rm Mpc}$, and $L_z=759.653~h^{-1}~{\rm Mpc}$, respectively.
(For the reasons given in footnote 6, this is approximately a third of the
volume to be surveyed by HETDEX.) Figure~\ref{fig:lognormal_galaxies} shows
the distribution of galaxies within the survey footprint of one realization.
The survey volume is no longer cuboidal in Cartesian coordinates, and the
$z$ direction is not parallel to the redshift direction due to curvature
of the sky. 
The galaxies within the survey footprint are approximately 2/3 of all galaxies
in the simulation box. From now on, we shall use the term ``geometry selection''
for galaxies within the survey footprint, and ``no selection'' for all galaxies
within the entire simulation box. The ``geometry selection'' galaxies are a
subset of the ``no selection'' galaxies.

To measure the power spectra from log-normal realizations, we use
the CIC density assignment to compute the local number density of
galaxies per density mesh, and then use eq.~\ref{eq:densitycontrast}
to compute the density contrast field, $\delta_g$. The number of
random samples used for computing $\bar{n}_g$ in eq.~\ref{eq:densitycontrast}
is approximately 2500 times that of galaxies, so that the Poisson
error in the estimate of $\bar{n}_g$ is negligible. Once the real-space
density contrast is constructed, it is Fourier transformed, and the
power spectrum computed by eq.~\ref{eq:Pg_hat}. To reduce the
sample variance we generate 1000 log-normal realizations, calculate
the averages of the power spectra of the ``no selection'' and
``geometry selection'' cases, and compute the ratio of the two,
$\bar P_{\rm geometry}(k)/\bar P_{\rm no}(k)$. We denote quantities
averaged over 1000 realizations with the over bar.

\begin{figure}[t]
\centering{
\includegraphics[width=1\textwidth]{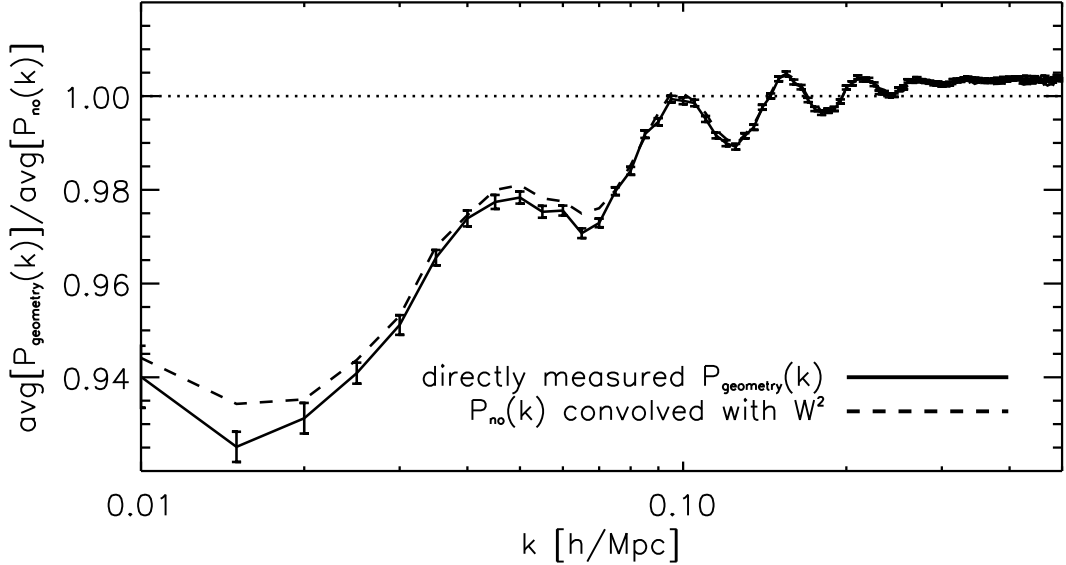}
}
\caption{Ratios of the average of the geometry selection power spectra,
 $\bar P_{\rm geometry}(k)$, to the average of the underlying
 (``no selection'') power spectra, $\bar P_{\rm no}(k)$, computed from
 1000 log-normal realizations (solid line; the error bars show the
 errors on the mean\protect\footnotemark). The dashed line shows
 the average of the underlying power spectra convolved with the window
 function squared, which agrees well with the direct measurement. Oscillatory
 features seen in $0.05\lesssim k\lesssim0.5~h~{\rm Mpc}^{-1}$ are not noise
 but real: they are produced by a smearing of BAO features due to the window
 function. See section~\ref{sec:bao} for details.
}
\label{fig:lognormal_pk_sel_geometry}
\end{figure}

\footnotetext{The errors on the mean is the scatter among the realizations
divided by the square root of the number of realizations.}

The solid line in figure~\ref{fig:lognormal_pk_sel_geometry} shows
this ratio, $\bar P_{\rm geometry}(k)/\bar P_{\rm no}(k)$. The ratio
is less than unity on large scales ($k\lesssim0.2~h~{\rm Mpc}^{-1}$),
whereas it approaches approximately unity on small scales.
\footnote{The ratio does not become exactly unity, as the peak of the window 
function for the geometry selection is broader than a delta function. As a
result, the window function-convolved power spectrum receives contributions
from adjacent Fourier modes, making the observed power spectrum on small
scales, where the underlying power spectrum declines steeply, slightly larger.}
This result is not surprising: the geometry selection has no effect on
the window function when the distance between galaxy pairs is smaller
than the curvature scale, and thus the geometry selection affects the
observed galaxy power spectrum only on large scales. This effect is
well known and presented in any galaxy surveys which are analyzed using
the usual Fourier transform technique.

It is straightforward to model this effect: convolve the squared Fourier
transform of the geometry selection function (see eq.~\ref{eq:Pconv_def}).
To compute the real-space selection function, we create
random particles in the simulation box, and select only the particles
within the survey footprint. The real-space selection function
is then proportional to the number density of the selected particles,
i.e., $W_{\rm num}({\bf r})\propto n_s({\bf r})$. The function,
$W_{\rm num}({\bf r})$, does not need to be normalized, as the
normalization factor is canceled by $W_{\rm sq}$ in the denominator
of eq.~\ref{eq:Pconv_def}.

The dashed line in figure~\ref{fig:lognormal_pk_sel_geometry} shows the average
of the ``no selection'' power spectra\footnote{The shot noise is subtracted
from the measured ``no selection'' power spectra before convolving them with
the window function squared.} measured from 1000 realizations convolved with
the squared window functions divided by the average of the ``no selection''
power spectra. The convolved power spectrum agrees well with the direct
measurement from the simulation shown by the solid line.

\subsection{Effect of sparse sampling}
\begin{figure}[t]
\centering{
\includegraphics[width=0.45\textwidth]{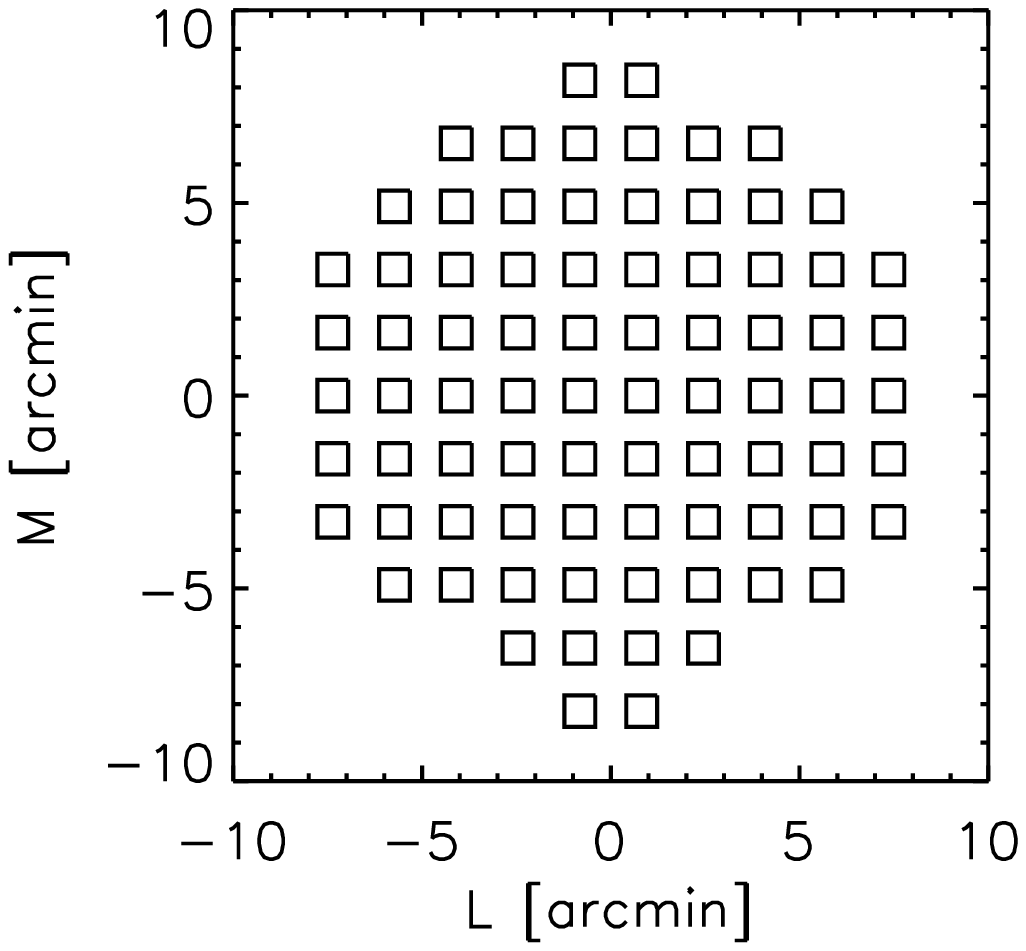}
\includegraphics[width=0.45\textwidth]{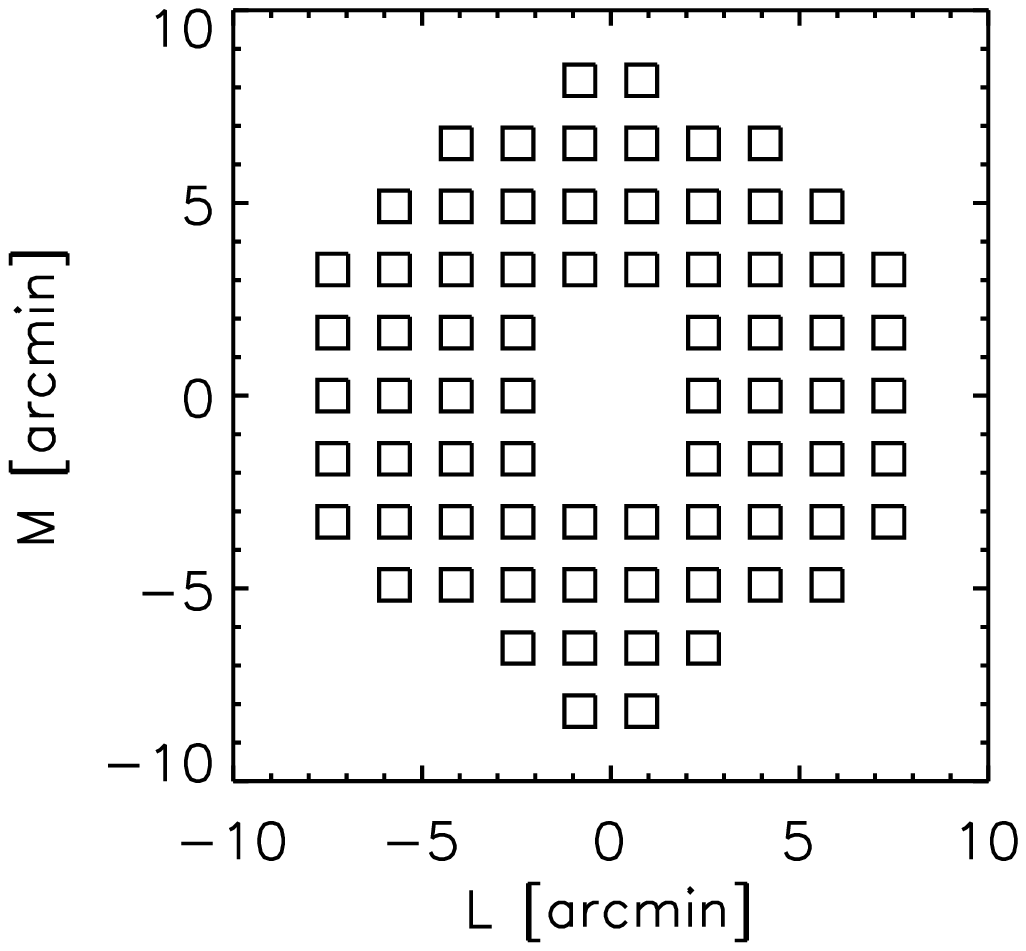}
}
\caption{
 (Left panel) Distribution of IFUs within a shot. Each square represents an IFU,
 and there are 80 IFUs. Six central IFUs in the left panel have been removed.
}
\label{fig:shot_pattern}
\end{figure}

\begin{figure}[t]
\centering{
\includegraphics[width=1\textwidth]{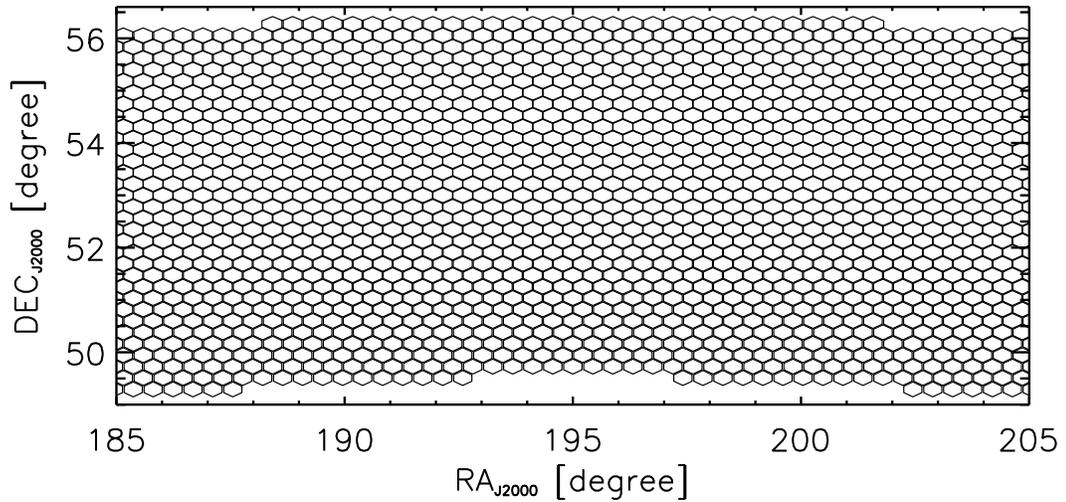}
}
\caption{Example shot positions in the sky. Each hexagon represents a
 shot location. 
 There are 4060 shots within the $34.1^{\circ}\times7.5^{\circ}$ survey area.
 As HET is more efficient for constant-declination observations,
 the shot positions are generated under the constant-declination
 requirement. There are no gaps or overlaps between shots at ${\rm
 DEC}=51.25^\circ$, whereas there are some gaps and overlaps  at lower and 
 higher declinations, respectively.
}
\label{fig:shot_position}
\end{figure}

We now apply the sparse sampling method to our simulations.
Figure~\ref{fig:shot_pattern} shows two example distributions
of IFUs within a single shot,\footnote{As VIRUS will come online in
stages, we use two different values (58 IFUs with a central hole for
the two-dimensional study and 74 IFUs with a central hole for the
three-dimensional study) to probe the possible effects. (Right panel)
Distribution of 74 IFUs within a shot.} and figure~\ref{fig:shot_position}
shows locations of shots in the northern sky. These locations resemble,
but are not the same as, the planned HETDEX survey footprint.

The shots are placed on constant declination rows, as HET is more efficient
for constant-declination surveys. Due to curvature of the celestial sphere,
it is impractical to design a realistic galaxy survey completely tiling the
survey area without any gaps or overlaps. 
In order to minimize excessive irregularities, gaps, or overlaps, we
first tile the shots without gaps or overlaps at ${\rm DEC}=51.25^{\circ}$. 
Then, for the adjacent (up and down) rows, the neighboring hexagon shots are
placed in phase to keep the shot locations regular.
We keep this procedure until the 
shots reach the boundary of the survey area. Because of curvature of the
celestial sphere, in higher declination rows some shots overlap and in lower
declination rows some shots have small gaps. The window function due to
these effects will be quantified later.

As we focus on the angular selection effect of sparse sampling in this paper,
we shall assume that all the IFUs have equal sensitivity at all wavelengths
for detecting galaxies. As this assumption will not hold in reality, there
will be a {\it radial} window function effect from the line-of-sight selection.
We have studied the radial window function of the HETDEX survey using the actual
sensitivity of IFUs measured from the HETDEX Pilot Survey
\cite{adams/etal:2011,blanc/etal:2011}, and found that the radial
selection function yields $|W(k_z)|^2/|W(0)|^2\lesssim 10^{-4}$, which
is much smaller than that of the angular selection. 

\begin{figure}[t]
\centering{
\includegraphics[width=1\textwidth]{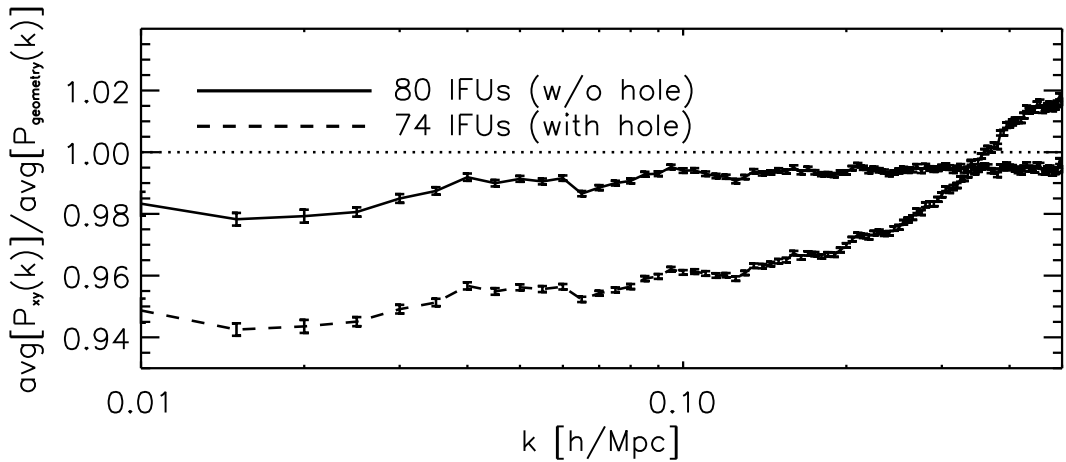}
}
\caption{Ratios of the averages of the 80-IFU (solid) and 74-IFU (dashed)
 selected power spectra to the average of the geometry selection power
 spectra. The solid line shows the window function effect due to gaps
 between shots toward lower declinations caused by curvature of the sky
 (see figure~\ref{fig:shot_position}), while the dashed line shows the
 additional window function effect caused by a hole in the middle of the
 focal plane (see the right panel of figure~\ref{fig:shot_pattern}).
 The error bars show the errors on the mean. The oscillatory features
 seen in this figure are not noise but real: they are caused by a smearing
 of BAO features due to the window function. See section~\ref{sec:bao}
 for details.
}
\label{fig:lognormal_pk_sel_xy}
\end{figure}

We select galaxies lying inside the IFUs, measure the power spectra from
1000 log-normal realizations, and average the results. To separate the
effect of sparse sampling from that of geometry selection, we divide the
averages of the 80-IFU and 74-IFU selected power spectra by the average
of the geometry selection power spectra (rather than by the underlying
power spectrum). Figure~\ref{fig:lognormal_pk_sel_xy} shows the ratios
of the averages of the 80-IFU (without a hole; solid line) and 74-IFU
(with a hole; dashed line) selected power spectra to the average of the
geometry selection power spectra. Ideally, if shot positions are regular
without any gaps and there is no large central hole (as in the
left panel of figure~\ref{fig:shot_pattern}), there is no window
function effect, as our one-dimensional and two-dimensional
studies have previously shown. However, because of curvature of the
celestial sphere, small gaps appear between shots at lower declinations
(see figure~\ref{fig:shot_position}), which yields a slight, $\sim2\%$
suppression on large scales. On the other hand, for the 74 IFUs shot
with a central hole, we observe a larger window function effect,
as expected from our two-dimensional study. 

The window function effect due to a central hole for the three-dimensional case
is smaller than that for the two-dimensional case, which is due to the smaller
fraction of the focal plane area occupied by the central hole (6 out of 80
here versus 6 out of 64 before).

\subsection{Baryon Acoustic Oscillation}
\label{sec:bao}
Thus far, we have focused our attention on the effects of sparse sampling and
curvature of the sky on the overall shape of the observed galaxy power spectrum,
and found a smooth suppression of the power on large scales. In this section,
we shall focus on the effects of sparse sampling and curvature of the sky
on sharper features in the power spectrum, i.e., BAO features.

To investigate how the window functions may affect BAOs, we extract BAOs from
the underlying power spectrum convolved with the window functions from various
cases such as the geometry selection and sparse sampling. We extract BAO
features as follows. We first fit the power spectra estimated from various cases
to a smooth power spectrum without BAO features as
\begin{equation}
 P_{\rm smooth}(k)=\sum_ic_iS_i(k) \ ,
\label{eq:Psmooth}
\end{equation}
where $S_i$ is the $i^{\rm th}$ cubic spline function, and
$c_i$ is a coefficient of $S_i$. We find $c_i$ by minimizing
$\chi^2=\sum_j[\langle\hat P_g(k_j)\rangle-\sum_ic_iS_i(k_j)]^2/\sigma^2(k_j)$,
where $\sigma^2(k_j)\equiv [\langle\hat P_g(k_j)\rangle+P_{\rm shot}]^2/N(k_j)$
and $N(k_j)$ is the number of independent Fourier modes in the $j^{\rm th}$ bin.
\footnote{The variance in the denominator of $\chi^2$, $\sigma^2$, is calculated
assuming that $\delta$ is a Gaussian field and $\chi^2$ is determined assuming
that adjacent Fourier modes are uncorrelated. Strictly speaking this is not a
good assumption as we are using log-normal realizations and the window function
correlates Fourier modes. However, this procedure still provides good estimates
of a smoothed power spectrum, i.e., the estimates are unbiased, but may not have
the minimum variance.} We set the fitting range to be
$0.005~h~{\rm Mpc}^{-1}\le k\le 0.4~h~{\rm Mpc}^{-1}$, and there are 12 parameters
to be fitted.\footnote{There are 10 break points in the fitting range:
$k=0.001~h~{\rm Mpc}^{-1}$ and $k=[0.02+i\times0.05]~h~{\rm Mpc}^{-1}$ with $i=0-8$.
The break points are set empirically.} We then extract BAOs as
\begin{equation}
 {\rm BAO}(k)\equiv\langle\hat{P}_g(k)\rangle-P_{\rm smooth}(k).
\label{eq:bao_def}
\end{equation}

\begin{figure}[t]
\centering{
\includegraphics[width=1\textwidth]{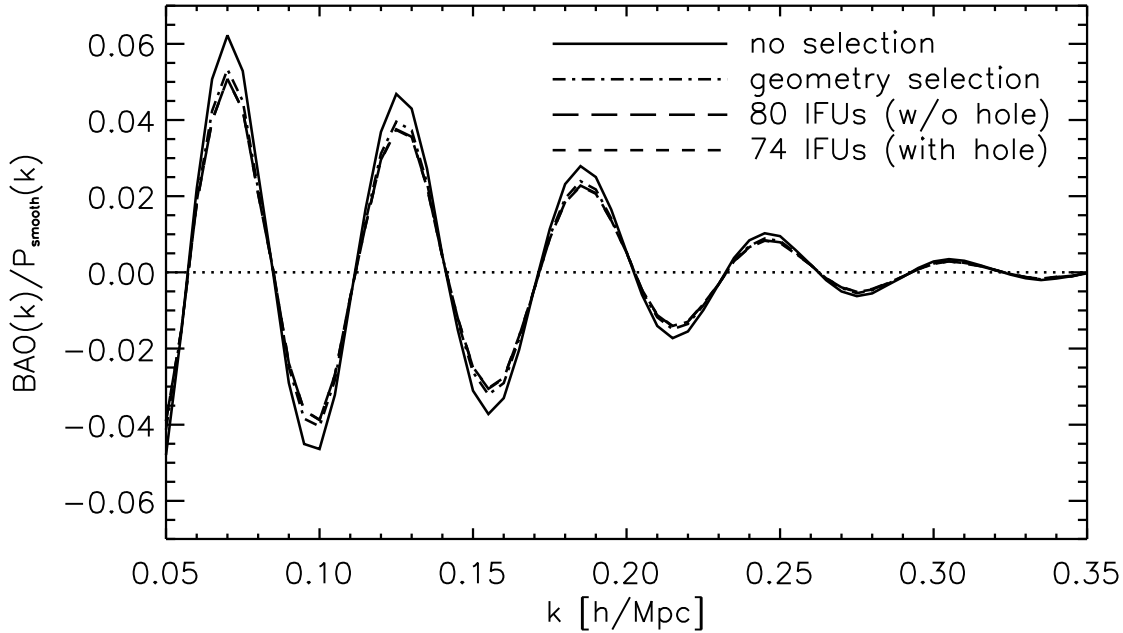}
}
\caption{Smearing of BAO features due to various window function effects.
 The solid line shows the underlying BAOs divided by the smooth power spectrum
 with no selection function. The dot-dashed line shows a smearing due to the
 application of Fourier transform to a spherical sky (``geometry selection'').
 The long-dashed line shows an additional smearing due to gaps between shots
 in lower declinations due to curvature of the sky (see figure~\ref{fig:shot_position}).
 The short-dashed line, which lies on top of the long-dashed line,
 demonstrates that a central hole in the focal plane (see the right panel of
 figure~\ref{fig:shot_pattern}) does not introduce a significant additional
 smearing of BAO features. The BAOs shown here are extracted from the convolution
 of the underlying power spectrum and the window functions.
}
\label{fig:bao_model}
\end{figure}

In figure~\ref{fig:bao_model}, we show the BAOs extracted from the
underlying power spectrum convolved with various window functions,
divided by the smooth power spectrum. We find that the BAO features
are smeared: while the phase of the BAO is unaffected, the amplitude
has decreased. The biggest smearing arises from the geometry selection
(the dot-dashed line; which has nothing to do with sparse sampling
but is caused by the application of Fourier transform to spherical sky),
which yields 12\% reduction in the amplitude of BAO features. Once again,
this effect is not new and is present in any survey results obtained
from the Fourier transform approach (see e.g.
\cite{tegmark/etal:2004,reid/etal:2010,cole/etal:2005,blake/etal:2010}).
One should in principle be able to remove this effect by using the spherical
Fourier-Bessel transform
\cite{heavens/taylor:1994,rassat/refregier:2011,leistedt/etal:2012}; 
we leave it as future work.

A smaller effect is produced by irregularities of the shot positions due
to curvature of the celestial sphere, which yield a further 5\%
reduction in the amplitude of BAOs (the long-dashed 
line). The central hole does not appear to introduce any additional smearing
(the short-dashed line, which lies on top of the long-dashed line) to the BAOs.

Smearing of BAOs occurs when BAOs are convolved with a smooth window function
with a broad width. The window function of the geometry selection does exactly
this. However, the window function of the sparse sampling is neither smooth
nor broad (see figure 1). Therefore, the sparse sampling with completely regular
separations causes {\it no} smearing of BAOs. We have seen this already from
figures 2, 5, and 6, where the ratios of the BAOs and the smooth power spectrum
on BAO scales do not have any structures. (I.e., both the BAOs and the smooth
power spectrum are suppressed by the same factor.) However, deviations from
regularity introduce a small additional smearing, as shown by the dashed line
in figure 12. 

\subsection{Constraint on the BAO peak position}
\label{sec:fitting}
\subsubsection{Finding the dilation parameter}
In the previous subsection, we have shown that the amplitude of BAOs
is smeared by the window function, mainly due to curvature of the sky.
Because of the smearing, the statistical power of the BAOs for cosmological
parameter estimations will degrade in proportion to the magnitude of smearing;
namely, if the amplitude of BAOs is reduced by 10\%, the parameter constraint
also degrades by 10\%, assuming that the uncertainty does not change. 
(If the signal-to-noise decreases by 10\% without changing the noise, it
is the signal that decreases by 10\%. Then, the uncertainty in the
parameters also increases by 10\%.) Moreover,
because of the window function, the adjacent Fourier modes of the power spectrum
may be correlated, and thus the parameter constraint degrades further.
In the following, we shall investigate how curvature of the sky and the sparse
sampling affect the constraint on the BAO peak position.

In order to study this issue, we estimate the so-called ``dilation parameter
,'' $\alpha$, of BAOs, i.e., ${\rm BAO}(k)\to {\rm BAO}(k/\alpha)$
\cite{anderson/etal:2012}, from 1000 log-normal simulations. 
We find $\alpha$ and the parameters characterizing the smooth component,
$c_i$, from each realization by minimizing
\begin{equation}
 \chi^2=\sum_{ij}\left[\hat{P}_g(k_i)-P_{\rm smooth}(k_i)-{\rm BAO}\left(\frac{k_i}{\alpha}\right)\right]
 \left[\hat{P}_g(k_j)-P_{\rm smooth}(k_j)-{\rm BAO}\left(\frac{k_j}{\alpha}\right)\right]C_{ij}^{-1}\ ,
\label{eq:chi2_corr}
\end{equation}
where $P_{\rm smooth}(k)$ is given in eq.~\ref{eq:Psmooth}, $C_{ij}$
is the covariance matrix computed from the power spectra of 1000 log-normal
realizations, and the BAO model here already includes the smearing due to
the window function effects, as shown in figure~\ref{fig:bao_model}.

\subsubsection{Structure of the covariance matrix}
\begin{figure}[t]
\centering{
\includegraphics[width=0.32\textwidth]{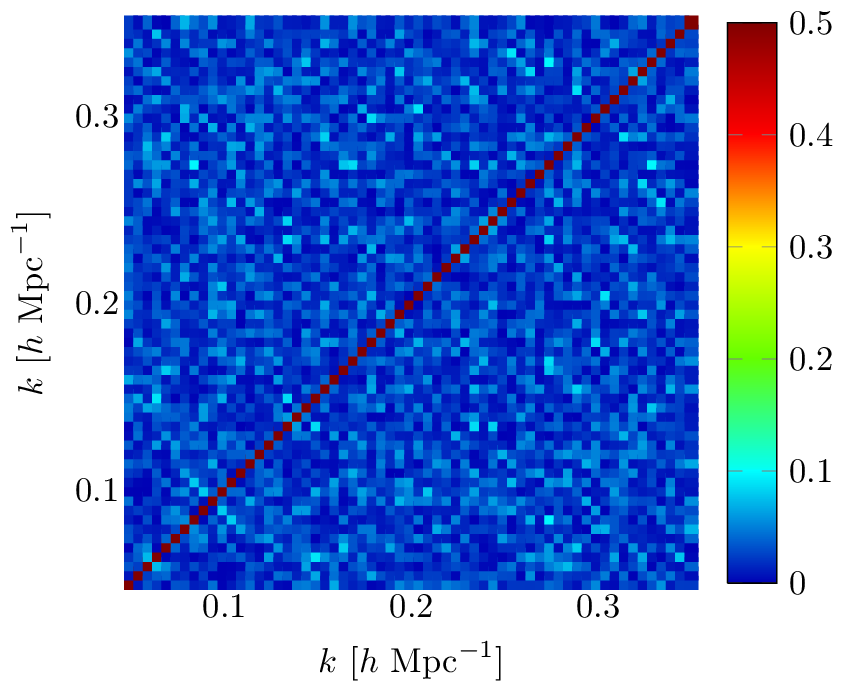}
\includegraphics[width=0.32\textwidth]{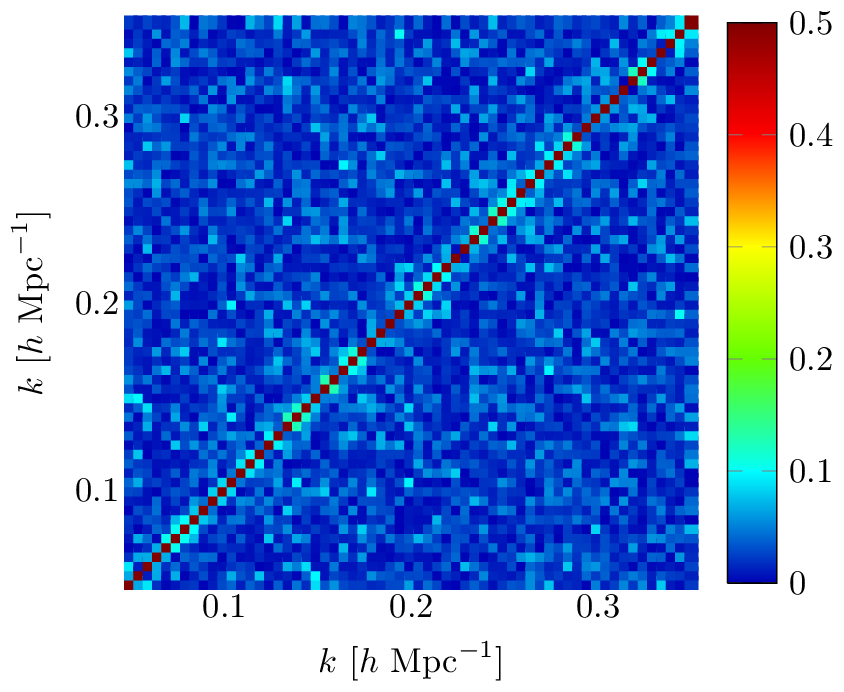}
\includegraphics[width=0.32\textwidth]{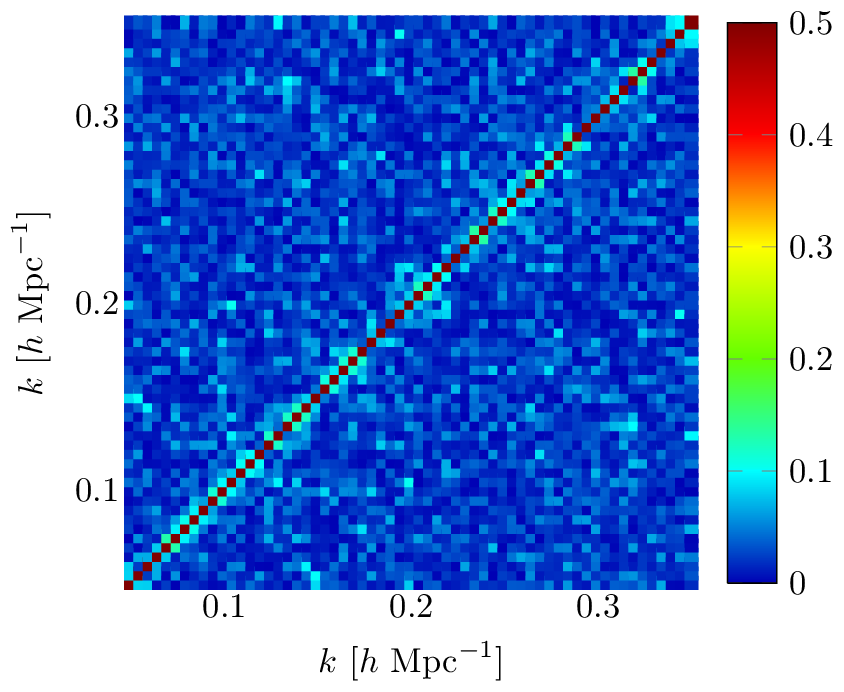}
\includegraphics[width=0.32\textwidth]{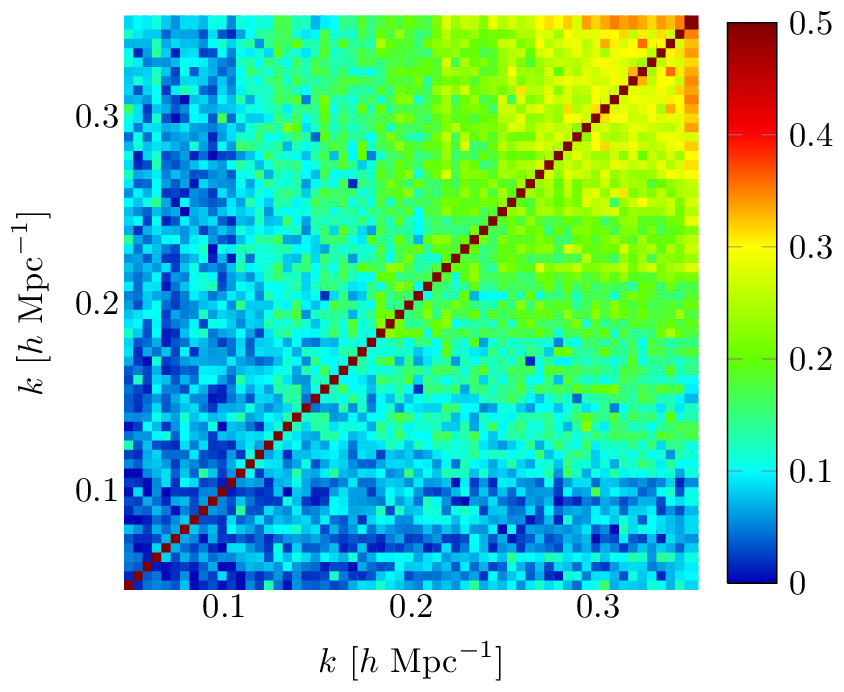}
\includegraphics[width=0.32\textwidth]{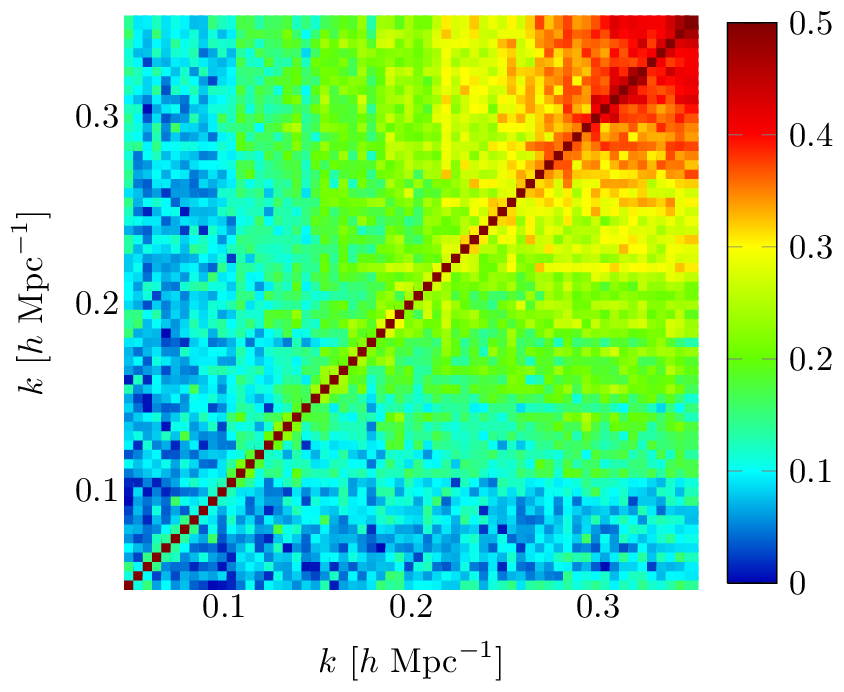}
\includegraphics[width=0.32\textwidth]{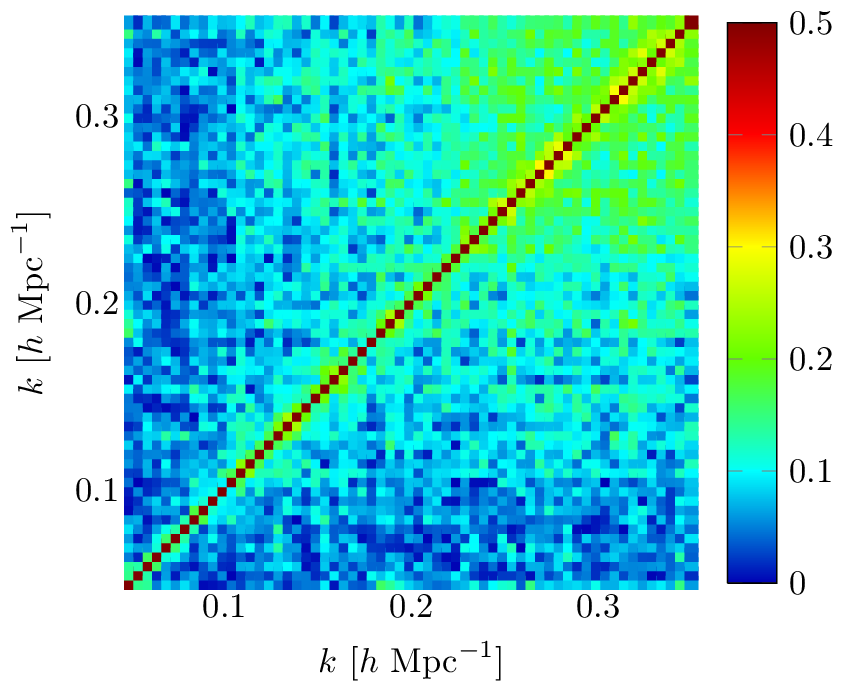}
}
\caption{Absolute values of the correlation coefficients of the
 covariance matrix, $|C_{ij}|/\sqrt{C_{ii}C_{jj}}$. We show the
 correlation coefficients for no selection (left panels), the
 geometry selection (middle panels), and the selection of 74 IFUs
 with a central hole (right panels). While we display the absolute
 values of the correlation coefficients, they are mostly positive.
 The diagonal elements are equal to unity, while the colors show
 values from 0 to 0.5. (Top panels) Correlation coefficients computed
 from 1000 Gaussian realizations, which do not include the effect of
 shot noise (see appendix~\ref{sec:gaussian}). These figures clearly
 show positive correlations of adjacent Fourier modes due to the
 window function effects. (Bottom panels) Correlation coefficients
 from 1000 log-normal realizations, which include the effect of shot noise.
 These figures show additional positive correlations due to the non-Gaussian
 nature of the underlying density fields. The bottom-right panel shows
 smaller correlation coefficients for off-diagonal elements, as the
 diagonal elements are enhanced by the larger shot noise without significantly
 increasing off-diagonal elements. (Recall that approximately 25\% of galaxies
 in the survey footprint are selected by 74 IFUs with a central hole,
 which results in much larger shot noise for the 74-IFU-selection.)
}
\label{fig:corr_matrix}
\end{figure}

It is necessary to compute $\chi^2$ using the full covariance matrix
(instead of only the diagonal elements of $C_{ij}$, i.e., the variance),
as the window function correlates adjacent Fourier modes.
In figure ~\ref{fig:corr_matrix}, we show the absolute values of
the correlation coefficient, $|C_{ij}|/\sqrt{C_{ii}C_{jj}}$,
for Gaussian (see appendix~\ref{sec:gaussian}) and log-normal
realizations. The wavenumbers shown here are $0.05\le k\le 0.35~h~{\rm Mpc^{-1}}$,
with increments of $0.005~h~{\rm Mpc}^{-1}$.\footnote{The fitting range
for the models ($0.005~h~{\rm Mpc}^{-1}\le k\le 0.4~h~{\rm Mpc}^{-1}$)
is chosen to be larger than that for the data, as we have to interpolate
for different values of $\alpha$, $0.9\lesssim\alpha\lesssim1.1$.}

The errors of the inverse covariance matrix $C^{-1}_{ij}$ are given by
$\langle\Delta(C^{-1})_{ij}^2\rangle=A(C^{-1})_{ij}^2+B[(C^{-1})_{ii}(C^{-1})_{jj}+(C^{-1})_{ij}^2]$
\cite{dodelson/schneider:2013}. In the Gaussian limit, $A=2/[(N_s-N_b-1)(N_s-N_b-4)]$
and $B=(N_s-N_b-2)/[(N_s-N_b-1)(N_s-N_b-4)]$, where $N_s$ and $N_b$ are the
number of realizations and bins, respectively. In our case, $N_s=1000$ and
$N_b=61$, yielding $A\simeq 2.28\times10^{-6}$ and $B\simeq 1.07\times10^{-3}$.
Thus, we estimate the errors on the inverse covariance matrix as
$\sqrt{\langle\Delta(C^{-1})_{ij}^2\rangle}\simeq 0.0327\sqrt{[(C^{-1})_{ii}(C^{-1})_{jj}+(C^{-1})_{ij}^2]}$.
Since the dominant components of the inverse covariance matrix are the
diagonal elements, the errors are bounded by
$0.0327\le\sqrt{\langle\Delta(C^{-1})_{ij}^2\rangle/[(C^{-1})_{ii}(C^{-1})_{jj}]}\le0.0462$.
As the inverse covariance matrix is dominated by the diagonal elements,
a few percent errors with respect to the diagonal elements of the off-diagonal
elements can be safely neglected. To test the convergence, we also run 5000
log-normal realizations. We find that the results of 5000 log-normal realizations
are quite similar to those of 1000 log-normal realizations, and the conclusions
of the paper are unchanged.

First, let us study the correlation coefficients computed from
Gaussian realizations. These realizations do not include the effect
of shot noise (see appendix~\ref{sec:gaussian}), and thus they show
purely the effects of window functions. For no selection (top left panel),
the off-diagonal elements of the correlation coefficients are negligible;
this is because the connected four-point function of $\delta({\bf k})$,
which produces non-zero off-diagonal elements of the covariance matrix,
vanishes for Gaussian random fields. (By definition the connected
four-point function is the four-point function minus the Gaussian
contribution.) However, this is true only when we
do not have window functions. In the presence of window functions,
the mode functions (i.e., $e^{i\mathbf{k}\cdot\mathbf{x}}$) are no longer
orthogonal, and thus adjacent Fourier modes become positively correlated.
We find this positive correlation for the geometry selection (top middle panel),
as well as for the sparse sampling of 74 IFUs with a hole (top right panel).
The correlations of adjacent Fourier modes are typically 10\%, and are
similar for both the geometry selection and the geometry plus IFU selection.
These results demonstrate that the primary source for the correlations
is the geometry selection, rather than the sparse sampling.

Next, let us study the correlation coefficients computed from log-normal
realizations (see appendix~\ref{sec:lognormal}), which include the effect
of shot noise. For no selection (bottom left panel), there are significant
positive correlations between Fourier modes due to the non-Gaussian nature
of the underlying density fields. For the geometry selection (bottom middle panel),
we find additional positive correlations for the adjacent Fourier modes 
(which is the same as those we found from Gaussian realizations),
as well as for a broad range of wavenumbers on small scales.

For the sparse sampling of 74 IFUs with a hole (bottom right panel),
the off-diagonal correlation coefficients become smaller than those of
the geometry selection.  This result is due to the larger shot noise:
there are much fewer galaxies selected by sparse sampling, and thus
the diagonal elements of the covariance matrix increase relative to
the off-diagonal elements. (The shot noise contributes only to the
diagonal elements of the covariance matrix.) As the correlation
coefficients are normalized to unity in the diagonal elements,
the off-diagonal elements become correspondingly smaller. The actual
(non-normalized) values of the off-diagonal elements of the covariance
matrix are slightly larger than those of the no-selection and geometry-
selection cases, in agreement with the results of Gaussian realizations.

\subsubsection{Uncertainty on $\alpha$: Fisher matrix versus direct fitting}
\label{sec:bao_alpha_fisher}
\begin{figure}[t]
\centering{
\includegraphics[width=1\textwidth]{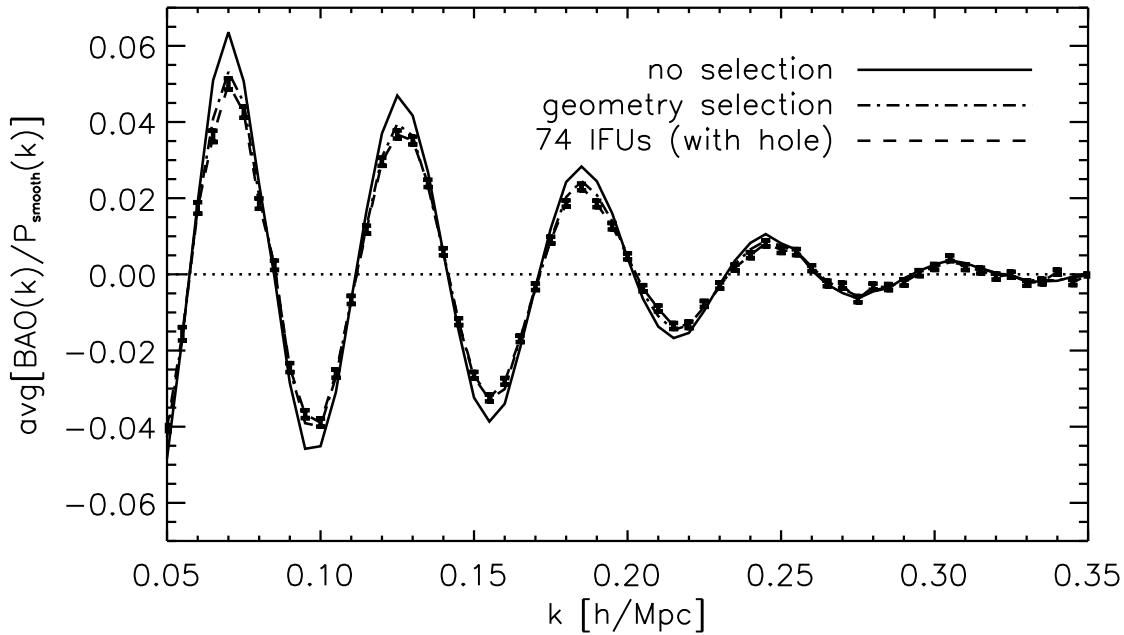}
}
\caption{Same as figure~\ref{fig:bao_model}, but for the average of BAOs
 extracted from 1000 realizations of our log-normal simulations.
 The error bars show the errors on the mean for 74 IFUs with a
 hole. The error bars for no- and geometry-selections are smaller.
}
\label{fig:bao_avg}
\end{figure}

With the covariance matrix computed, we now minimize $\chi^2$
given in eq.~\ref{eq:chi2_corr} to find the values for $\alpha$ from 1000
realizations, and compute the 1-$\sigma$ uncertainty in $\alpha$.
We set the fitting range to be $0.05~h~{\rm Mpc}^{-1}\le k\le0.35~h~{\rm
Mpc}^{-1}$, which contains most of the BAO features.
There are 61 data points, and 11 parameters (10 for $c_i$ and 1 for
$\alpha$) to be fitted in total. We use the Levenberg-Marquardt method
(\cite{numerical_recipes}) to find the minimum of $\chi^2$.

The averages of BAOs extracted from 1000 realizations of our log-normal
simulations for various window functions are shown in figure~\ref{fig:bao_avg},
which are in agreement with the models in figure~\ref{fig:bao_model}.

Using 1000 log-normal realizations, we find that the average of
$\alpha$ from 1000 realizations is unity to within the uncertainty of
simulations (i.e., the standard deviation divided by $\sqrt{1000}$),
indicating that our method yields an unbiased estimate of $\alpha$.
The 1-$\sigma$ uncertainties in $\alpha$ we find are 0.92\%, 1.17\%,
and 1.95\% for no selection, geometry selection, and sparse sampling
of 74 IFUs with a central hole, respectively.\footnote{If the errors on
computation of the inverse covariance matrix are taken into account, 
variance on the fitted parameters would increase by a factor of
$1+N_b/N_s=1.061$, i.e., the 1-$\sigma$ uncertainty increases by 1.03.}  

To check whether log-normal realizations agree with expectations,
let us compare these values with the expectations from the simplest
treatment of galaxy surveys that is widely used by the cosmology
community. Ignoring the off-diagonal elements of the covariance matrix
or the effect of window functions on the diagonal elements of the
covariance matrix, the uncertainty in $\alpha$ is given by
\begin{equation}
\frac1{({\rm Err}[\alpha])^2}
 =V_{\rm survey}\int_{k_{min}}^{k_{max}}\frac{d^3k}{2(2\pi)^3}
 \frac{1}{[\bar P_g(k)+1/\bar{n}_g]^2}\left[\frac{\partial{\rm BAO}(k)}
 {\partial{\rm ln}k}\right]^2 \ .
\label{eq:conv_fisher_matrix}
\end{equation}
Here, $\bar P_g(k)$ is the average power spectrum of 1000 log-normal realizations
with corresponding window functions,
and ${\rm BAO}(k)$ is given by the BAOs extracted from the underlying power
spectrum convolved with window functions, as shown in figure~\ref{fig:bao_model}. 

In this formula, we need the survey volume, $V_{\rm survey}$, and the galaxy
number density, $\bar{n}_g$. The meaning of these quantities is clear for
the no-selection and geometry-selection cases: as all galaxies within the
cuboid simulation box are observed, the volume of no-selection case is
$1.018~h^{-3}~{\rm Gpc}^3$. For the geometry selection, only galaxies lying
within the survey footprint are observed, and thus the volume is
$0.722~h^{-3}~{\rm Gpc}^3$ (255.75 square degrees in the sky and $1.9<z<2.5$).
As the observed regions are contiguous, the number density is identical
for both the no-selection and geometry-selection cases. We have
$\bar{n}_g=2.95\times10^{-3}~h^3~{\rm Mpc}^{-3}$. (There are 3.003 million and
2.130 million galaxies for no-selection and geometry-selection cases,
respectively.)  

What about in the sparse sampling case?
The results we have presented in this paper so far suggest that,
provided that the wavenumber of interest is sufficiently smaller
than the wavenumber corresponding to the separation between IFUs
($k\simeq 3.14~h~{\rm Mpc}^{-1}$), the sparse sampling approach can
yield the same results as the survey which has the volume within
the survey footprint (i.e., the outermost boundary of the survey),
and the number density of galaxies which is the total number of
observed galaxies divided by the volume of the footprint.
We thus take the survey volume to be the volume of the footprint,
$V_{\rm survey}=0.722~h^{-3}~{\rm Gpc}^3$, and the galaxy number
density for the selection by 74 IFUs with a hole to be
$\bar{n}_g=0.667\times10^{-3}~h^3~{\rm Mpc}^{-3}$. (There are 0.482 million
observed galaxies for the selection by 74 IFUs with a central hole.)

Inserting these values into eq.~\ref{eq:conv_fisher_matrix}, we find the
expected uncertainties in $\alpha$ of 0.86\%, 1.21\%, and 1.90\% for
no selection, geometry selection, and sparse sampling of 74 IFUs with
a central hole, respectively.\footnote{If one uses the sum of the sub-volumes
for sparse sampling, then $V_{\rm survey}=0.108~h^{-3}~{\rm Gpc}^3$ and
$\bar{n}_g=2.95\times10^{-3}~h^3~{\rm Mpc}^{-3}$, and the expected
uncertainty in $\alpha$ becomes 3.32\%, which is roughly 60\% larger than
that measured in our log-normal realizations.} These numbers are in good
agreement with 0.92\%, 1.17\%, and 1.95\% we find from the direct fitting of
log-normal realizations. 

The Fisher matrix given by eq.~\ref{eq:conv_fisher_matrix} does
not include the off-diagonal elements of the covariance matrix
or the fact that we simultaneously fit 11 parameters. To include these
effects, we generalize eq.~\ref{eq:conv_fisher_matrix} to \cite{tegmark:1997}
\begin{equation}
 F_{ij}=\sum_{m,n}\frac{\partial\langle\hat{P}_g(k_m)\rangle}{\partial\theta_i}
 \frac{\partial\langle\hat{P}_g(k_n)\rangle}{\partial\theta_j}C^{-1}_{mn} \ ,
\label{eq:original_fisher_alpha}
\end{equation}
where $F_{ij}$ is the Fisher matrix, and $\theta_i$ denotes the
$i^{\rm th}$ parameter. As there are 11 parameters, the dimension of the
Fisher matrix is $11\times11$, and we set $\theta_0\equiv\alpha$ and
$\theta_i\equiv c_i$. The 1-$\sigma$ uncertainty in $\alpha$ is then given
by $\sqrt{(F^{-1})_{00}}$. We find ${\rm Err}[\alpha]=$ 0.90\%, 1.13\%,
and 1.76\% for no selection, the geometry selection, and the sparse sampling
of 74 IFUs with a central hole, respectively.

From these results, we conclude that the uncertainty in $\alpha$
achieved by the sparse sampling is 
comparable to that of a galaxy survey with the survey volume of the footprint,
and the number density given by the total number of observed galaxies
within the footprint divided by the volume of the footprint.

\subsubsection{Significance of BAO detection with sparse sampling}
\label{sec:bao_detection}
In section~\ref{sec:bao_alpha_fisher}, we presented the averages of the
extracted BAOs from 1000 log-normal realizations. In this subsection, we
shall show the extracted BAOs from six log-normal realizations as
examples, and discuss the significance of BAO detection with sparse sampling.

\begin{figure}[t]
\centering{
\includegraphics[width=1\textwidth]{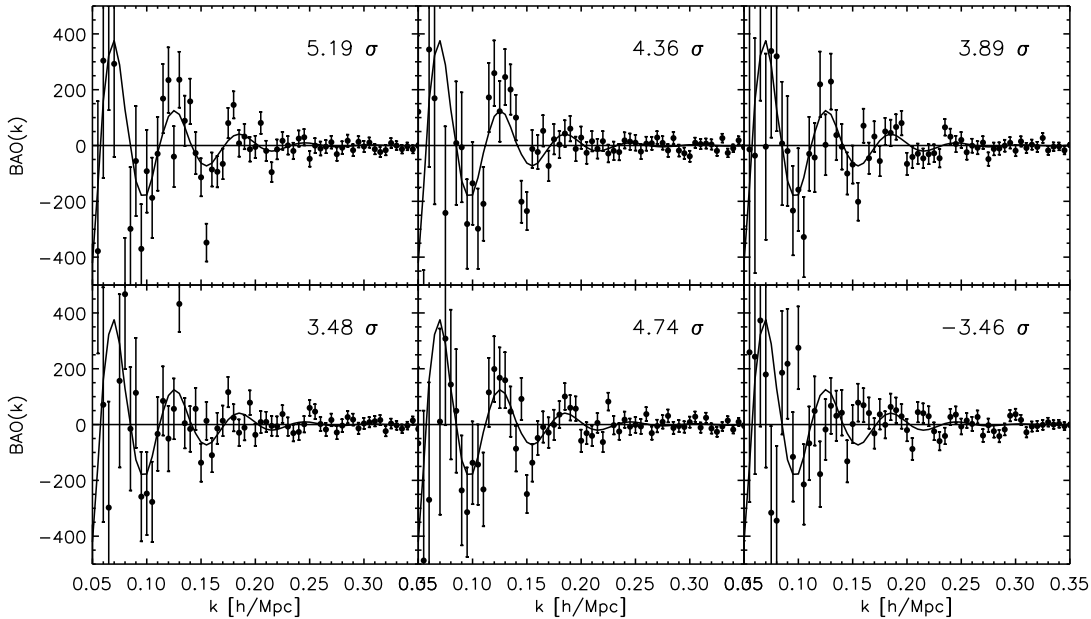}
}
\caption{
 Extracted BAOs from six randomly-chosen log-normal realizations with
 sparse sampling. The ``$\sigma$'' labels in the upper right of each
 panel show ``significance of BAOs detection,'' calculated as
 $\sqrt{\chi^2_{\rm null}-\chi^2}$, where $\chi^2$ and $\chi^2_{\rm null}$
 are the minimum $\chi^2$ values with and without BAOs included in the
 model. In the bottom-right panel, we find $\chi^2>\chi^2_{\rm null}$
 (i.e., we do not detect BAOs) and thus we show $-\sqrt{\chi^2-\chi^2_{\rm
 null}}$.  
}
\label{fig:bao_6_realizations}
\end{figure}

In figure~\ref{fig:bao_6_realizations}, we display the extracted BAOs,
$\hat P_g(k)-P_{\rm smooth}(k)$, from six randomly-chosen log-normal
realizations with sparse sampling. The error bars are given by the
square root of diagonal elements of the covariance matrix directly
measured from our log-normal realizations. The values of $\sigma$
is the significance of BAO detection.

We quantify the significance of BAO detection for each realization by
$S/N=\sqrt{\chi^2_{\rm null}-\chi^2}$, where $\chi^2$ is the minimum $\chi^2$
value obtained by minimizing eq.~\ref{eq:chi2_corr}, and $\chi^2_{\rm null}$
is the minimum $\chi^2$ value obtained by minimizing
\begin{equation}
 \chi^2_{\rm smooth}=\sum_{ij}[\hat{P}_g(k_i)-P_{\rm smooth}(k_i)]
 [\hat{P}_g(k_j)-P_{\rm smooth}(k_j)]C_{ij}^{-1}\ ,
\end{equation}
which is the minimum $\chi^2$ value for a model without BAO features.

\begin{figure}[t]
\centering{
\includegraphics[width=1\textwidth]{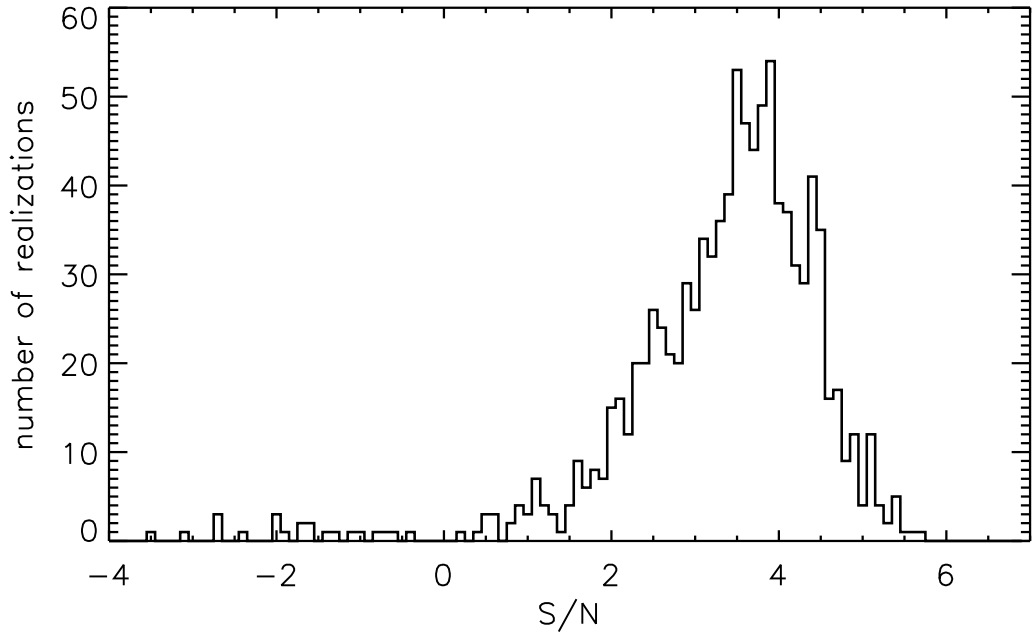}
}
\caption{Distribution of the significance of BAO detection (``the number
 of sigmas'') from 1000 log-normal realizations with sparse sampling of
 74 IFUs with a central hole.
}
\label{fig:bao_detection_sig}
\end{figure}

In figure~\ref{fig:bao_detection_sig}, we present the distribution of
the significance of BAO detection from our 1000 log-normal realizations
with sparse sampling of 74 IFUs with a central hole. While we find
$\chi^2_{\rm null}>\chi^2$ (i.e., BAOs provide a better fit) from most
of the realizations, there are also $\chi^2_{\rm null}<\chi^2$ from sixteen
unlucky realizations out of 1000. In such a case we use
$S/N=-\sqrt{\chi^2-\chi^2_{\rm null}}$, which takes on negative values
and means that we do not detect any BAOs. Our study shows that, for the
survey volume and the number of galaxies we use in these simulations,
BAOs are detected typically at the $3.5\sigma$ levels.\footnote{For the
reasons described in footnote 6, this volume corresponds to about a
third of the total volume that would be surveyed by HETDEX.}

\subsection{Constraint on the amplitude of the power spectrum}
\label{sec:amp}
In the previous subsection, we have focused on measuring the peak
position of the BAOs. To show that sparse sampling works not only
for extracting the peak position of the BAOs but also for extracting
the other parameters, we shall measure the {\it amplitude} of the
power spectrum from 1000 log-normal realizations in this subsection.

First, we compute the average power spectra of 1000 log-normal
realizations, $\bar P_g(k)$,  for no selection, geometry selection,
and sparse sampling (74 IFUs with a central hole) and use them as
the models. Then, we fit the amplitude by minimizing
\begin{equation}
 \chi^2=\sum_{ij}[\hat{P}_g(k_i)-A\bar P_g(k_i)][\hat{P}_g(k_j)-A\bar P_g(k_j)]C_{ij}^{-1},
\label{eq:chi2_amp}
\end{equation}
with respect to $A$; the solution is
\begin{equation}
 A=\frac{\sum_{ij}C^{-1}_{ij}[\hat P_g(k_i)\bar P_g(k_j)+\hat P_g(k_j)\bar P_g(k_i)]}
 {2\sum_{ij}C^{-1}_{ij}\bar P_g(k_i)\bar P_g(k_j)}.
\end{equation}
Finally, by setting the fitting range to be from
$k_{min}=0.05~h~{\rm Mpc}^{-1}$ to $k_{max}=0.35~h~{\rm Mpc}^{-1}$,
we compute the fractional uncertainties of $A$: ${\rm Err}[A]/A=0.71\%$,
0.87\%, and 1.03\% for no selection, geometry selection, and sparse sampling
of 74 IFUs with a central hole, respectively.

The expected uncertainty of the amplitude of the power spectrum from
the Fisher matrix is given by
\begin{equation}
\frac{1}{({\rm Err}[A])^2}
 =V_{\rm survey}\int_{k_{min}}^{k_{max}}\frac{d^3k}{2(2\pi)^3}
 \Bigg[\frac{\bar P_g(k)}{\bar P_g(k)+1/\bar{n}_g}\Bigg]^2,
\label{eq:amp_fisher_matrix}
\end{equation}
where we have used that $\partial\ln\bar P_g(k)/\partial\ln A=1$. Using
the values of the survey volume and the mean number density, which are the
same as those in the previous subsection, we find the expected uncertainties
of ${\rm Err}[A]/A=0.22\%$, 0.26\%, and 0.45\%, for no selection, geometry
selection, and sparse sampling of 74 IFUs with a central hole, respectively.

The expected uncertainties from the Fisher matrix are too small compared
to the direct fitting, as eq.~\ref{eq:amp_fisher_matrix} ignores
off-diagonal elements of the covariance matrix, which are mainly due to 
non-Gaussianity of the log-normal density fields.\footnote{To prove
that the off-diagonal elements are mainly due to non-Gaussianity of the
density fields, we fit the amplitude of the power spectrum extracted from
1000 {\it Gaussian} realizations, and find ${\rm Err}[A]/A=$0.17\%, 0.19\%,
and 0.20\% for no selection, geometry selection, and sparse sampling of
74 IFUs with a central hole, respectively. As there is no shot noise in our
Gaussian realizations, we set $1/\bar{n}_g=0$ in eq.~\ref{eq:amp_fisher_matrix},
and find the expected 1-$\sigma$ uncertainties of the amplitude of the
power spectrum to be 0.16\%, 0.20\%, and 0.20\% for no selection, geometry
selection, and sparse sampling of 74 IFUs with a central hole, respectively.
The agreement we find from our Gaussian realizations shows that the disagreement
we find from our log-normal (hence non-Gaussian) realizations is due to
non-Gaussianity of the log-normal density fields.} To include the off-diagonal
elements, we modify eq.~\ref{eq:original_fisher_alpha} as
\begin{equation}
 F_{AA}=\sum_{m,n}\frac{1}{\bar P_g(k_m)}\frac{1}{\bar P_g(k_n)}C^{-1}_{mn},
\label{eq:original_fisher_amp}
\end{equation}
and the 1-$\sigma$ uncertainty in $A$ is then given by $1/\sqrt{F_{AA}}$.
We find ${\rm Err}[A]/A= 0.71$\%, 0.87\%, and 1.03\% for no selection, the
geometry selection, and the sparse sampling of 74 IFUs with a central hole,
respectively. The results are in good agreement with the direct fitting.

If we use only the diagonal elements of the covariance matrix in
eq.~\ref{eq:original_fisher_amp}, i.e., $C^{-1}_{mn}=0$ for $m\neq n$
and $C^{-1}_{mn}$ is the inverse of the variance of the power spectrum for $m=n$,
we find ${\rm Err}[A]/A=$0.24\%, 0.26\%, and 0.41\%, for no selection, the
geometry selection, and the sparse sampling of 74 IFUs with a central hole,
respectively. These values are in good agreement with those from
eq.~\ref{eq:amp_fisher_matrix}, again indicating that the survey volume
for sparse sampling should be the survey footprint.

We can summarize our finding for sparse sampling as follows.
Let the volume of the survey footprint be $V_{\rm survey}$,
the sum of the volumes of the observed regions be $V_{\rm observe}$,
and the total number of observed galaxies be $N_{\rm gal}$.
For wavenumbers smaller than the wavenumber
corresponding to the
separations between observed regions, the survey volume is $V_{\rm survey}$,
and the fractional uncertainty of the power spectrum is proportional to
$V_{\rm survey}^{-1/2}[P_g(k)+N_{\rm gal}/V_{\rm survey}]$.
On the other hand, for wavenumbers larger than the wavenumber corresponding
to the separations 
between observed regions, the survey volume is $V_{\rm observe}$, and the
fractional uncertainty of the power spectrum is proportional to
$V_{\rm observe}^{-1/2}[P_g(k)+N_{\rm gal}/V_{\rm observe}]$.

\section{Two-point correlation function with sparse sampling}
\label{sec:2pcf}

The basic reason why the power spectrum is biased by the window function
is that computation of the power spectrum requires an estimate of
density fields, which we then Fourier transform.
Computation of the the two-point correlation function in configuration
space, on the other hand, does not require an estimate of density
fields: we can simply count the number of pairs and compare it to the
expectation from random pairs distributed over observed regions. This
process automatically corrects for the effect of sparse sampling. In
this section, we shall show that the two-point correlation function in
configuration space is not affected by sparse sampling.

Assuming that particle 1 and 2 are at locations ${\bf r}_1$ and ${\bf
r}_2$ from us, we
define the separation $r=|{\bf r}_1-{\bf r}_2|$ and the line-of-sight
vector as $({\bf r}_1+{\bf r}_2)/2$. We then compute the correlation
function using the Landy-Szalay (LS) estimator \cite{landy/szalay:1993} as
\begin{equation}
 \xi(r,\mu)= \frac{N_R(N_R-1)/2}{N_D(N_D-1)/2}\frac{DD(r,\mu)}{RR(r,\mu)}
 -\frac{N_R(N_R-1)/2}{N_RN_D}\frac{2DR(r,\mu)}{RR(r,\mu)}+1 \ ,
\label{eq:2pcf_def}
\end{equation}
where $DD(r,\mu)$, $DR(r,\mu)$, and $RR(r,\mu)$ are the numbers of galaxy-galaxy
pairs, galaxy-random pairs, and random-random pairs, respectively, and
$\mu$ is the cosine between the line-of-sight and the tangential
directions (i.e., ${\bf r}_1-{\bf r}_2$). Finally we spherically average the
correlation function as $\xi(r)=\int_0^1d\mu~\xi(r,\mu)$. 
Note that the random particles have the same selection function as the galaxies.
In eq.~\ref{eq:2pcf_def}, $N_D$ ad $N_R$ are the numbers of galaxies and random
particles, respectively, to normalize the numbers of pairs. Namely,
there are $N_R(N_R-1)/2$ random-random pairs, $N_D(N_D-1)/2$
galaxy-galaxy pairs, and $N_RN_D$ galaxy-random pairs. We use 20 million
random particles for no selection, and approximately 15 million random
particles (after sparse sampling) for sparse sampling of 74 IFUs with a
central hole.

Unlike the power spectrum, we expect the correlation function not to be
affected by the window function effect. Dividing the number of galaxy-galaxy
and galaxy-random pairs by the number of the random-random pairs with random
particles having the same selection as the galaxies, we automatically
correct for the selection effect. Of course, as less galaxies
are observed 
due to sparse sampling, the uncertainties of the correlation function would
increase compared to the no-selection case. Thus, the only effect of
sparse sampling on the correlation function is the increasing uncertainties.

\begin{figure}[t]
\centering{
\includegraphics[width=1\textwidth]{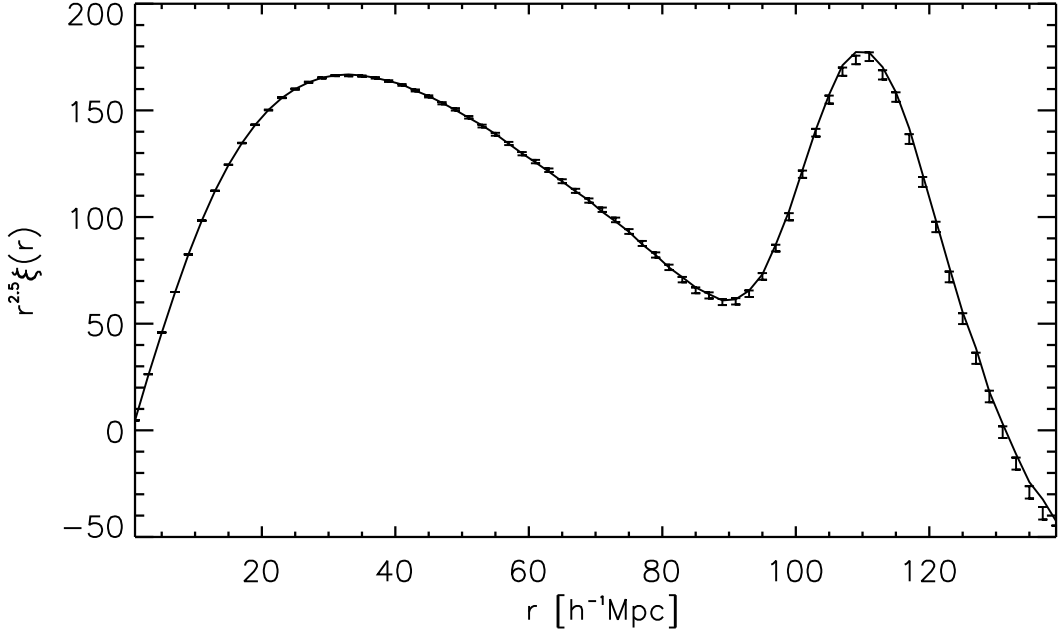}
}
\caption{Average two-point correlation function computed from 1000 log-normal
 realizations with no selection (solid line) and sparse sampling of 74 IFUs
 with a central hole. The error bars show errors on the mean for 74~IFUs with a hole. 
}
\label{fig:2pcf}
\end{figure}

Figure~\ref{fig:2pcf} shows the average two-point correlation function measured
from 1000 log-normal realizations with no selection (solid line) and sparse sampling
of 74 IFUs with a central hole (dots with error bars). We multiply the correlation
function by $r^{2.5}$ to emphasize the BAO bump at $r\simeq 110~ h^{-1}~$~Mpc.
We find that the solid line (no selection) and the data points with error bars
(measurement with sparse sampling) are statistically consistent. (Note that the
error bars of the correlation functionare strongly correlated.)

\begin{figure}[t]
\centering{
\includegraphics[width=1\textwidth]{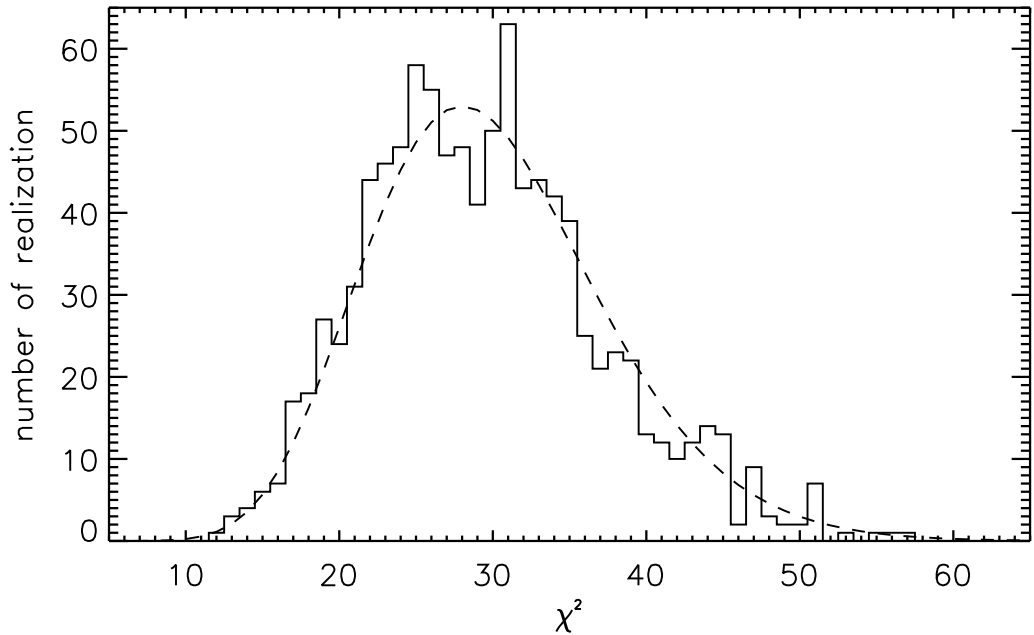}
}
\caption{$\chi^2$ of the correlation functions of 1000 log-normal realizations
 with sparse sampling of 74 IFUs with a central hole with respect to the mean
 correlation function of the no-selection case. The dashed line is the
 $\chi^2$-distribution  with 30 degrees of freedom.
}
\label{fig:2pcf_chi2}
\end{figure}

To quantify the agreement between them, we compute the goodness-of-fit
for the correlation functions of 1000 log-normal realizations with
sparse sampling of 74 IFUs with a central hole with respect to
the mean correlation function of the no-selection case. We compute
\begin{equation}
 \chi^2=\sum_{ij}[\xi_g(r_i)-\bar{\xi}(r_i)](C^{-1})_{ij}[\xi_g(r_j)-\bar{\xi}(r_j)],
\end{equation}
where $C_{ij}$ is the covariance matrix of the correlation function,
$\xi_g(r)$ is the correlation function of one realization with sparse
sampling of 74 IFUs with a central hole, and $\bar{\xi}(r)$ is the
average of 1000 correlation functions with no selection. Setting the range to 
be $70~h^{-1}~{\rm Mpc}\le r\le130~h^{-1}~{\rm Mpc}$, we have 30 bins.
The histogram of $\chi^2$ is shown in \ref{fig:2pcf_chi2}. We find
that the  histogram of $\chi^2$ follows a $\chi^2$-distribution with 30
degrees of freedom. This result shows that $\xi_g(r_i)$ is a fair
representation of the underlying $\bar{\xi}(r_i)$.

\section{Conclusion}
\label{sec:conclusion}
In this paper, we have shown how to perform a galaxy redshift survey with
sparse sampling, when the goal is to measure the galaxy power spectrum.
Sparse sampling can be an effective approach in situation where a large focal
plane is subdivided into many smaller and sparsely distributed
apertures, e.g., IFUs. 

Our basic findings are straightforward and can be summarized as follows:

Suppose that one wishes to cover a survey volume of $V_{\rm survey}$.
It is not necessary to observe galaxies within the entirety of $V_{\rm survey}$; 
rather, one may divide $V_{\rm survey}$ into many sub-volumes, and
observe galaxies within these sub-volumes. The sub-volumes are separated
by some distance, $r$. For efficient sparse sampling, $r$ should be comparable to  
the linear size of each sub-volume (e.g., twice the linear size of each
sub-volume). However, it is important to set $r$ smaller than the spatial
scale corresponding to the wavenumbers of interest. In other words, when
calculating a density field from observed galaxy locations, the size of the
density mesh must be chosen such that the Nyquist frequency, $k_{\rm Nyq}$,
of the mesh is lower than the frequency corresponding to the separation between
observed sub-volumes, i.e., $k_{\rm Nyq}<2\pi/r$. Also, the distribution of
sub-volumes must be chosen to be as regular as possible: random positions
would suppress the large-scale power the most. When the regular sparse sampling
is achieved, the survey is as powerful as a survey which covers $V_{\rm survey}$
and has the number density of galaxies given by the total number of galaxies
observed within sub-volumes divided by $V_{\rm survey}$. Regular sparse sampling
in the flat sky yields no window function effect on the observed galaxy power
spectrum, and the BAO features are preserved with no smearing.

However, there is one drawback, which is the number density of galaxies.
As the number density of galaxies determines the uncertainty of the measured
galaxy power spectrum at a given wavenumber, it is important to make sure that
the number density of galaxies satisfies the constraint, $\bar{n}_gP(k_{\rm max})\gtrsim 1$,
where $k_{\rm max}$ is the maximum wavenumber below which the power spectrum is measured.
As sparse sampling yields far fewer galaxies within $V_{\rm survey}$ than the filled survey,
a longer integration per shot to collect more galaxies within each sub-volume is
necessary. Therefore, there is an interesting trade-off between the effectiveness
of sparse sampling and the steepness of a luminosity function of a target galaxy
population. If the luminosity function is steep, a modest increase in the integration
time per shot yields many more galaxies, and thus sparse sampling can be quite efficient.
If, however, the luminosity function is not steep enough, one must
integrate every shot
for too long to obtain the sufficient number of galaxies, and thus it may be more
efficient to fill $V_{\rm survey}$ contiguously by doing a shallower survey. For
HETDEX, the size of the telescope (10~m) helps: the current estimate
suggests that 
it takes only 20 minutes per shot to reach $L^*$ of the luminosity
function of Lyman-$\alpha$ emitting galaxies and to collect enough
Lyman-$\alpha$ emitting galaxies at $1.9\le z\le 3.5$.
Shorter duration shots would incur too much observational overhead, and
thus sparse sampling is an efficient way to increase the  
survey volume. For example, $\bar{n}_gP_g(k)$ is expected to be 6.86 and 0.42 at
$k=0.05~h~{\rm Mpc}^{-1}$ and $0.35~h~{\rm Mpc}^{-1}$, respectively.

In addition, we have explored the effects of several real-world issues, including:
\begin{itemize}
 \item [1.] {\bf Tolerable randomness}. The distribution of observed regions
       cannot be completely regular because, e.g., locations of bright stars
       and large nearby galaxies must be avoided. This adds some degree of
       randomness to the distribution of the observed regions in the sky. We
       find that this is not an issue as long as a typical displacement is
       less than 10\% of the separation between the observed regions. 
 \item [2.] {\bf Gaps and rotation}. Our study clearly shows that one
       must make the best effort to avoid visible (i.e., bigger than
       the size of the density mesh) gaps between the observed regions.
       If gaps are unavoidable due to, e.g., curvature of the celestial
       sphere, the window function is calculable and correctable.
       Rotation of orientations of shots does not introduce significant
       effects on the window function.
 \item [3.] {\bf Holes on the focal plane}. Holes on the focal plane
       due to, e.g., a need to place different instruments, introduce
       a window function effect whose magnitude is given by the fraction
       of the focal plane area occupied by holes. Holes should be avoided
       in general; however, if absolutely necessary, the window function
       is calculable and correctable. Fortunately, holes occupying $\sim 10$\%
       of the focal plane do not appear to introduce significant additional
       smearing of the BAO features in the power spectrum.  
\end{itemize}

The largest window function effect that we have identified in our
three-dimensional study, which is done within the context of HETDEX,
has nothing to do with sparse sampling, and it is due to
an application of Fourier transform to the spherical sky (``geometry
selection''). This effect should be eliminated by using the spherical
Fourier-Bessel expansion.

We find that the geometry selection smears out the amplitude of BAO by
$\sim 10$\%, and sparse sampling combined with the real-world effects
such as gaps toward lower declinations and holes in the middle of the
focal plane yields an additional smearing at the level of $\sim 5$\%.
Once the smearing is taken into account, the uncertainty on the distance
scale measured by BAO, $\alpha$, agrees with the expected uncertainty
assuming the survey volume of $V_{\rm survey}$ and the number density
given by the total number of galaxies observed within the sub-volumes
divided by $V_{\rm survey}$, to within 10\%. The remaining differences
in the uncertainty in $\alpha$ relative to the expectation largely arise
from the geometry selection, which should diminish if we use the spherical
Fourier-Bessel transform. The confirmation of the latter statement is
subject to future work.

Finally, we have shown that the two-point correlation function as
computed by pair counting is not affected by sparse sampling.

Sparse sampling provides an efficient and affordable solution to a
problem of executing large galaxy redshift surveys covering more than
$10~{\rm Gpc}^3$ of survey volume. While our study is presented within
the context of HETDEX, the sparse-sampling method itself is general and
can be applied to other galaxy surveys. 

Finally, we have focused solely on the effect of
sparse sampling on measurements of the galaxy power spectrum. How sparse
sampling affects higher-order correlations such as the bispectrum and
trispectrum is left as future work.

\acknowledgments
We thank the members of the HETDEX collaboration for comments on the
paper, as well as for continuous encouragement and support.
This material is based in part upon work supported by NASA grants NNX08AM29G
and NNX08AL43G, and by NSF grant AST-0807649.
We acknowledge the Texas Advanced Computing Center (TACC;
{\sf http://www.tacc.utexas.edu}) at The University of Texas at Austin
for providing high-performance computing resources that have contributed
to the research results reported within this paper.
DJ acknowledges DoE SC-0008108 and NASA NNX12AE86G.
HETDEX is run by the University of Texas at Austin
McDonald Observatory and Department of Astronomy with participation from
the Ludwig-Maximilians-Universit\"at M\"unchen,
Max-Planck-Institut f\"ur Extraterrestriche-Physik (MPE),
Leibniz-Institut f\"ur Astrophysik Potsdam (AIP),
Texas A\&M University, Pennsylvania State University,
Institut f\"ur Astrophysik G\"ottingen, University of Oxford,
and Max-Planck-Institut f\"ur Astrophysik (MPA).
In addition to Institutional support, HETDEX is funded by
the National Science Foundation (grant AST-0926815),
the State of Texas, the US Air Force (AFRL FA9451-04-2-0355),
by the Texas Norman Hackerman Advanced Research Program under
grants 003658-0005-2006 and 003658-0295-2007, and by
generous support from private individuals and foundations.


\appendix
\section{Gaussian perturbations to regularly-spaced sparse sampling}
\label{sec:perturbation}
In this section, we derive the expected window function
for the (one-dimensional) regularly-spaced sparse sampling,
with Gaussian perturbations to observed positions.

In this context, the expectation
value of the window function squared is given by
\begin{equation}
 \langle\left|W(k)\right|^2\rangle=
 \left[d\ {\rm sinc}\left(\frac{kd}{2}\right)\right]^2
 \sum_{a=1}^N\sum_{b=1}^Ne^{-ik\bar{x}_a}e^{ik\bar{x}_b}
 \langle e^{-ik\epsilon_a}e^{ik\epsilon_b}\rangle.
\label{eq:Wk_expect}
\end{equation}
To proceed, we first Taylor-expand $\langle e^{-ik\epsilon_a}e^{ik\epsilon_b}\rangle$ as
\begin{equation}
 \langle e^{-ik\epsilon_a}e^{ik\epsilon_b}\rangle=
 \langle\sum_{c=0}^{\infty}\frac{(-ik)^c}{c!}\epsilon_a^c
 \sum_{d=0}^{\infty}\frac{(ik)^d}{d!}\epsilon_b^d\rangle
 =\sum_{c=0}^{\infty}\sum_{d=0}^{\infty}\frac{(-1)^c(ik)^{c+d}}{c!d!}
 \langle\epsilon_a^c\epsilon_b^d\rangle \ .
\label{eq:exp_epsilon_expectation}
\end{equation}
Then, we apply $\langle\epsilon_a\rangle=0$,
$\langle\epsilon_a\epsilon_b\rangle=\sigma_{\epsilon}^2\delta_{ab}$,
and Wick's theorem to simplify eq.~\ref{eq:exp_epsilon_expectation}.
There are five situations.
\begin{itemize}
\item[1.  ] $a=b$ and $c+d=$odd
\begin{equation}
 \langle\epsilon_a^c\epsilon_b^d\rangle=
 \langle\epsilon_a^{c+d}\rangle=0
\label{eq:situaion_1}
\end{equation}

\item[2.  ] $a=b$ and $c+d=0$
\begin{equation}
 \langle\epsilon_a^c\epsilon_b^d\rangle=1
\label{eq:situation_2}
\end{equation}

\item[3.  ] $a=b$ and $c+d=$even
\begin{equation}
 \langle\epsilon_a^c\epsilon_b^d\rangle=
 \langle\epsilon_a^{c+d}\rangle=(c+d-1)!!\sigma_{\epsilon}^{c+d}
\label{eq:situation_3}
\end{equation}

\item[4.  ] $a\neq b$ and ($c=$odd or $d=$odd)
\begin{equation}
 \langle\epsilon_a^c\epsilon_b^d\rangle=0
\label{eq:situation_4}
\end{equation}

\item[5.  ] $a\neq b$ and ($c=$even and $d=$even)
\begin{equation}
 \langle\epsilon_l^a\epsilon_m^b\rangle=(a-1)!!(b-1)!!\sigma_{\epsilon}^{a+b}
\label{eq:situation_5}
\end{equation}

\end{itemize}
Inserting the above results into eq.~\ref{eq:Wk_expect}, we obtain
\begin{eqnarray}
 &&\sum_{a=1}^N\sum_{b=1}^Ne^{-ik\bar{x}_a}e^{ik\bar{x}_b}
 \langle e^{-ik\epsilon_a}e^{ik\epsilon_b}\rangle\nonumber\\
 &=&\sum_{a=1}^N\left[1+\sum_{c=1}^{\infty}
 \sum_{d=0}^{2c}\frac{(-1)^d(ik)^{2c}}{d!(2c-d)!}(2c-1)!!\sigma_{\epsilon}^{2c}\right]\nonumber\\
 &+&\sum_{a\neq b}e^{-ik(\bar{x}_a-\bar{x}_b)}\left[
 1+2\sum_{c=1}^{\infty}\frac{(ik)^{2c}}{(2c)!}(2c-1)!!\sigma_{\epsilon}^{2c}
 +\sum_{c=1}^{\infty}\sum_{d=1}^{\infty}\frac{(ik)^{2c+2d}}{(2c)!(2d)!}
 (2c-1)!!(2d-1)!!\sigma_{\epsilon}^{2c+2d}\right] \ .\nonumber\\
\label{eq:expand_Wk_expect}
\end{eqnarray}
In eq.~\ref{eq:expand_Wk_expect}, the terms can be computed as
\begin{eqnarray}
 &&\sum_{c=1}^{\infty}\sum_{d=0}^{2c}
 \frac{(-1)^d(ik)^{2c}}{d!(2c-d)!}(2c-1)!!\sigma_{\epsilon}^{2c}
 =\sum_{c=1}^{\infty}\sum_{d=0}^{2c}\frac{(-1)^d(ik)^{2c}}{d!(2c-d)!}
 \frac{(2c)!}{2^cc!}\sigma_{\epsilon}^{2c}\nonumber\\
 &=&\sum_{c=1}^{\infty}\sum_{d=0}^{2c}\frac{(-1)^d(ik)^{2c}}{2^cc!}
 \frac{(2c)!}{d!(2c-d)!}\sigma_{\epsilon}^{2c}
 =\sum_{c=1}^{\infty}\frac{(ik)^{2c}\sigma_{\epsilon}^{2c}}{2^cc!}
 \sum_{d=0}^{2c}(-1)^d\binom{2c}{d}=0 \ ,
\label{eq:term_1}
\end{eqnarray}
\begin{equation}
 \sum_{c=1}^{\infty}\frac{(ik)^{2c}}{(2c)!}(2c-1)!!\sigma_{\epsilon}^{2c}
 =\sum_{c=1}^{\infty}\frac{(-k^2\sigma_{\epsilon}^2)^c}{c!2^c}
 =\sum_{c=0}^{\infty}\frac{(-k^2\sigma_{\epsilon}^2)^c}{c!2^c}-1
 =e^{-k^2\sigma_{\epsilon}^2/2}-1\ ,
\label{eq:term_2}
\end{equation}
and
\begin{eqnarray}
 &&\sum_{c=1}^{\infty}\sum_{d=1}^{\infty}\frac{(ik)^{2c+2d}}{(2c)!(2d)!}
 (2c-1)!!(2d-1)!!\sigma_{\epsilon}^{2c+2d}\nonumber\\
 &=&\sum_{c=1}^{\infty}\frac{(-k^2\sigma_{\epsilon}^2)^c(2c-1)!!}{(2c)!}
 \sum_{d=1}^{\infty}\frac{(-k^2\sigma_{\epsilon}^2)^j(2d-1)!!}{(2d)!}
 =(e^{-k^2\sigma_{\epsilon}^2/2}-1)^2 \ .
\label{eq:term_3}
\end{eqnarray}
Finally, we obtain
\begin{eqnarray}
 \langle\left|W(k)\right|^2\rangle&=&
 \left[d\ {\rm sinc}\left(\frac{kd}{2}\right)\right]^2\left[\sum_{a=1}^N 1+
 e^{-k^2\sigma_{\epsilon}^2}\sum_{a\neq b}e^{-ik(\bar{x}_a-\bar{x}_b)}\right]\nonumber\\
 &=&\left[d\ {\rm sinc}\left(\frac{kd}{2}\right)\right]^2\left[N+
 2e^{-k^2\sigma_{\epsilon}^2}\sum_{a>b}e^{-ik(\bar{x}_a-\bar{x}_b)}\right] \ .
\label{eq:Wk_expect_final}
\end{eqnarray}

\section{Log-normal simulation}
\label{sec:lognormal}
The log-normal simulation is a relatively inexpensive method to generate
realizations of non-linear density fluctuations. While the power spectrum
measured from these realizations on small scales may deviate from the
underlying power spectrum, the extracted BAOs are in good agreement.
In this appendix, we describe our log-normal simulation. 

Ref.~\cite{kayo/taruya/suto:2001} shows that the density contrast of
matter computed from N-body simulations follows a log-normal distribution.
This result motivates our generating random realizations of density fields
drawn from a log-normal distribution. 

The log-normal density contrast is defined as
\begin{equation}
 G({\bf x})=\ln[\delta({\bf x})+1]-\langle\ln[\delta({\bf x})+1]\rangle \ ,
\label{eq:G_def}
\end{equation}
where $G({\bf x})$ follows Gaussian statistics. The density contrast
can be written as $\delta({\bf x})=Ae^{G({\bf x})}-1$, where
$A\equiv\exp\left[\langle\ln[\delta({\bf x})+1]\rangle\right]$ is the
normalization factor. Since the ensemble average of the density contrast is 0,
we find
\begin{eqnarray}
 \frac{1}{A}=\langle e^{G}\rangle
 =\sum_{n=0}^{\infty}\frac{\langle G^n\rangle}{n!}
 =\sum_{k=0}^{\infty}\frac{(2k-1)!!}{(2k)!}\sigma_G^{2k}
 =\sum_{k=0}^{\infty}\frac{1}{k!}\left(\frac{\sigma_G^2}{2}\right)^k
 =\exp\left(\frac{\sigma_G^2}{2}\right) \ ,
\label{eq:A_value}
\end{eqnarray}
where $\sigma_G^2=\langle G^2\rangle$ is the variance of the Gaussian field.
Thus, combining eq.~\ref{eq:G_def} and eq.~\ref{eq:A_value},
one can rewrite the density contrast as
$\delta({\bf x})=e^{-\sigma_G^2/2}\exp[G({\bf x})]-1$.

The two-point correlation function of the log-normal density contrast is
\begin{equation}
 \xi({\bf x})=\langle\delta({\bf x_1})\delta({\bf x_2})\rangle
 =e^{-\sigma_G^2}\langle e^{G({\bf x_1})}e^{G({\bf x_2})}\rangle-1 \ .
\label{eq:xi_def}
\end{equation}
As $G({\bf x})$ is a Gaussian random field,
the correlation function of the exponent is
\begin{equation}
 \langle e^{G({\bf x_1})}e^{G({\bf x_2})}\rangle
 =e^{\sigma_G^2+\xi_G({\bf x})} \ ,
\label{eq:eG_corr_def}
\end{equation}
where $\xi_G({\bf x})=\langle G({\bf x_1})G({\bf x_2})\rangle$
is the two-point correlation function of the Gaussian random field.
Finally, one can relate the correlation function
of the log-normal density contrast and the Gaussian random field as
$\xi({\bf x})=e^{\xi_G({\bf x})}-1$ or
$\xi_G({\bf x})=\ln\left[\xi({\bf x})+1\right]$.

To generate log-normal realizations, we inverse-Fourier-transform
the underlying power spectrum, $P({\bf k})$, to obtain the two-point
correlation function, $\xi({\bf x})$; calculate the two-point correlation
function of the Gaussian random field, $\xi_G({\bf x})$; and Fourier-
transform $\xi_G({\bf x})$ back to find $P_G({\bf k})$. Then, we generate
the Gaussian random field in Fourier space as
\begin{equation}
 G({\bf k})=\sqrt{0.5P_G({\bf k})}\theta \ ,
\label{eq:Gk_def}
\end{equation}
where $\theta$ is a complex Gaussian random variable with unit variance.
The factor of $0.5$ in eq.~\ref{eq:Gk_def} is due to the reality condition.
We then inverse-Fourier-transform $G({\bf k})$ to real space and
calculate $\sigma_G^2$.
As we have seen already, the relation between  the log-normal density
contrast and the Gaussian random field is given by $\delta({\bf
x})=e^{-\sigma_G^2/2}e^{G({\bf x})}-1$, which assures $\delta({\bf x})\ge-1$.
Finally, the number of the galaxies within a given mesh is
\begin{equation}
 N({\bf x})=\bar{n}\left[1+\delta({\bf x})\right]V_{\rm mesh} \ ,
\label{eq:N_def}
\end{equation}
where $\bar{n}=N_{\rm total}/V_{\rm survey}$.
As the number of galaxies in a mesh is an integer,
we generate an integer value drawn from a Poisson distribution with the
mean given by eq.~\ref{eq:N_def}.

\begin{figure}[t]
\centering{
\includegraphics[width=0.48\textwidth]{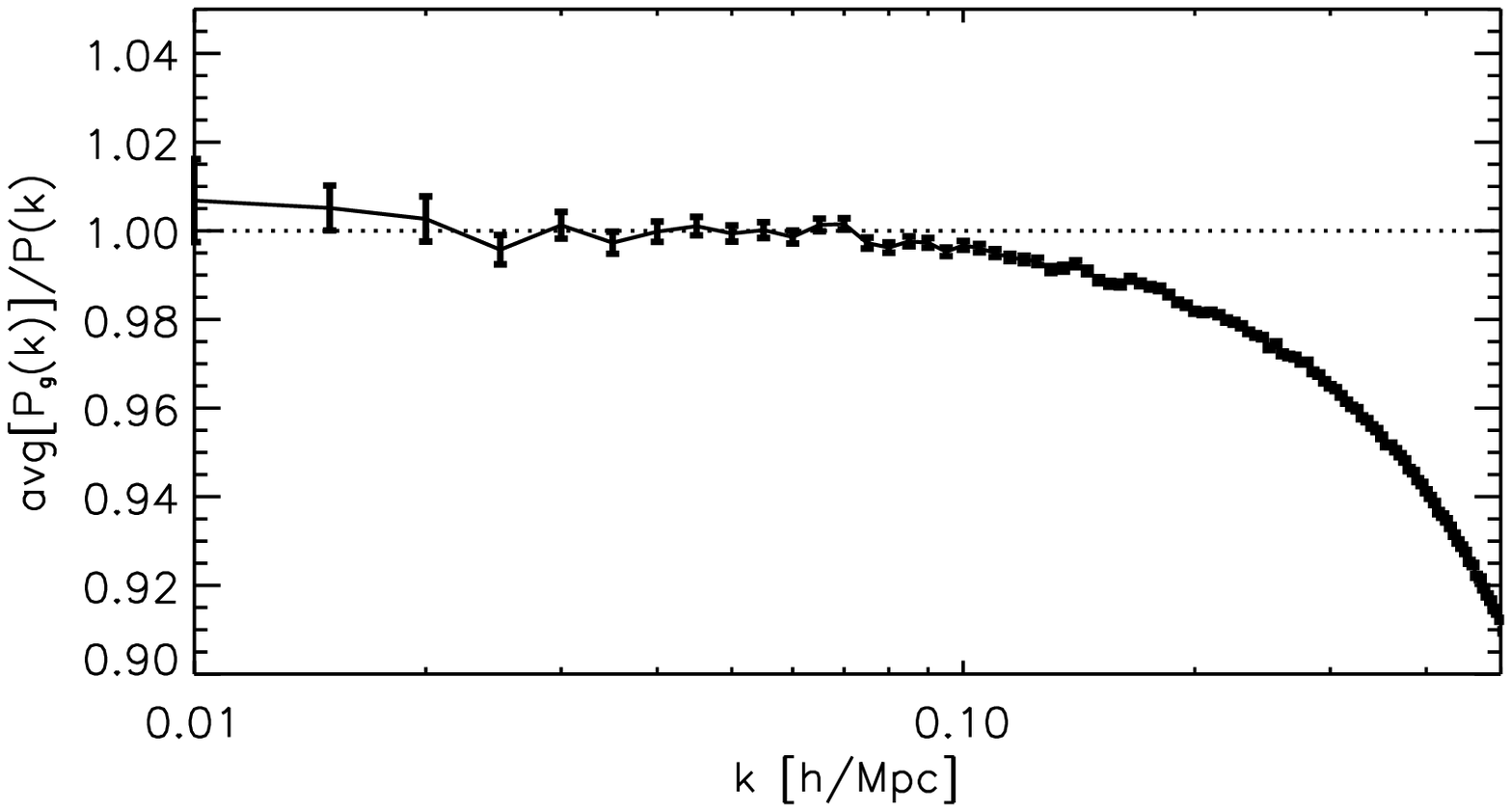}
\includegraphics[width=0.48\textwidth]{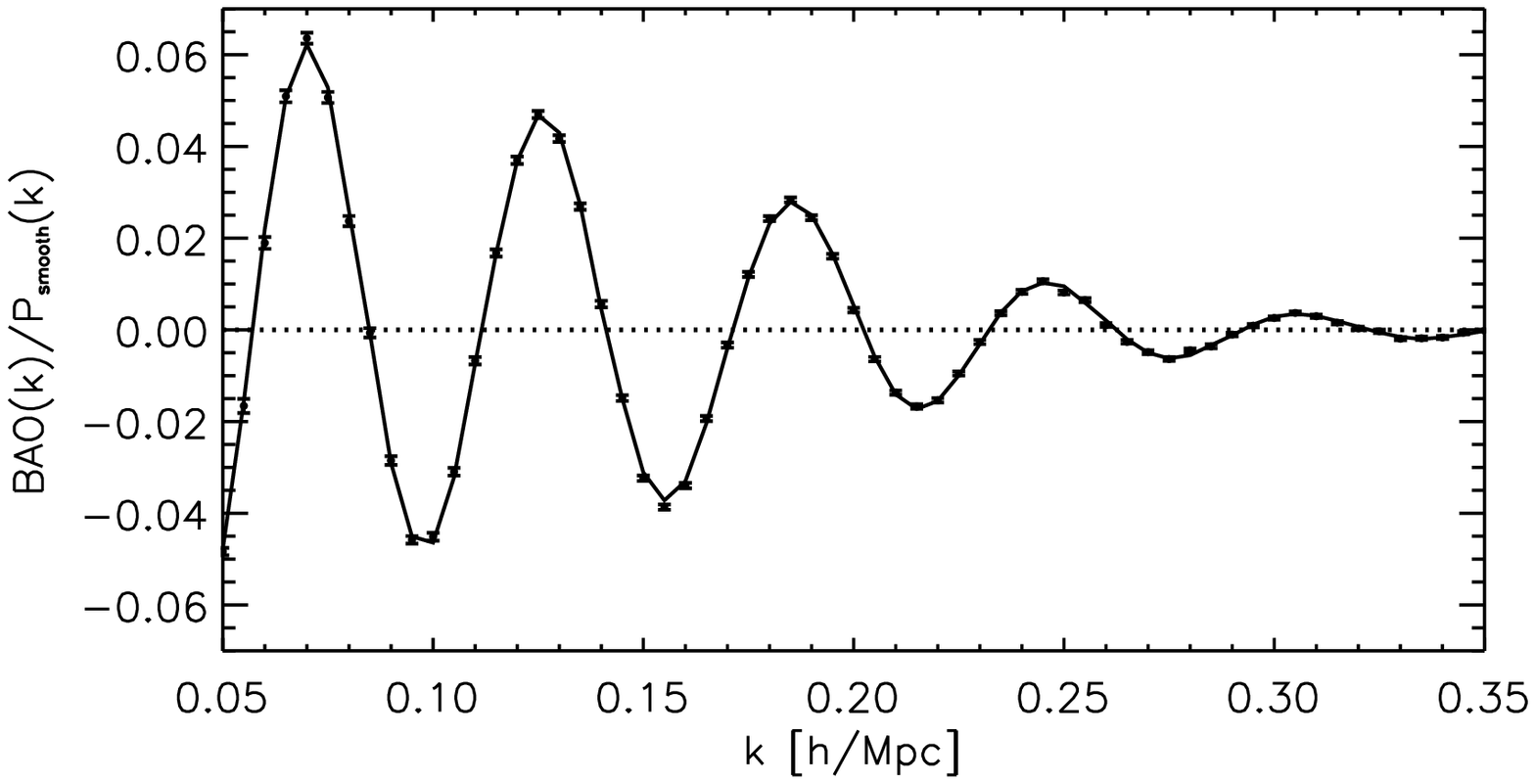}
}
\caption{
 (Left panel) Ratio of the average of 1000 galaxy power spectra measured
 from log-normal realizations to the underlying power spectrum.
 (Right panel) BAOs extracted from the underlying power spectrum (solid line)
 and the average of 1000 BAOs extracted from log-normal realizations.
 The error bars show the errors on the mean.
}
\label{fig:lognormal_pk_sel_no}
\end{figure}

We generate log-normal realizations using the input galaxy power spectrum
at $z=2.2$ computed from the third order perturbation theory with non-linear
bias \cite{jeong/komatsu:2006,jeong/komatsu:2009}. The bias parameters are
$b_1=2.2$, $b_2=0.671$, and $P_0=72.13~h^{-3}~{\rm Mpc}^3$, and the number
density of galaxies is $\bar{n}=2.95\times 10^{-3}~h^3~{\rm Mpc^{-3}}$.
The cosmological parameters are $\Omega_m=0.26$ and $\Omega_{\Lambda}=0.74$.
Following these procedures, we generate 1000 log-normal realizations,
compute power spectra, and then extract BAOs.

In the left panel of figure ~\ref{fig:lognormal_pk_sel_no}, we show the
ratio of the average of the 1000 measured power spectra to the underlying
power spectrum. The average measured power spectrum from log-normal realizations
is in excellent agreement with the underlying power spectrum at
$k\lesssim0.1~h~{\rm Mpc}^{-1}$; however, the measured spectrum is suppressed
relative to the input for $k\gtrsim0.1~h~{\rm Mpc}^{-1}$. This effect is due to
the resolution of the density mesh used for generating Gaussian random fields
(the Nyquist frequency is $k_{\rm Nyq}=3.12~h~{\rm Mpc}^{-1}$ for
figure~\ref{fig:lognormal_pk_sel_no}). The agreement should extend to higher
$k$ if we use higher resolution.

Nevertheless, we find that this resolution is sufficient for accurately
recovering the BAOs. The right panel of figure~\ref{fig:lognormal_pk_sel_no}
shows the BAO of the underlying power spectrum (solid line) and the average
of 1000 BAOs extracted from 1000 log-normal realizations (dots with error bars).
The method for extracting BAOs is described in section~\ref{sec:bao}. 
We find an excellent agreement between the two. 

\section{Gaussian simulation of density fields with window functions but
 without shot noise}
\label{sec:gaussian}
While log-normal realizations are useful for generating a semi-realistic
distribution of galaxies, normal, Gaussian realizations
are also useful for isolating the effect of window functions.

A Gaussian density field is generated from 
\begin{equation}
 \delta({\bf k})=\sqrt{0.5P({\bf k})}\theta \ ,
\label{eq:gaussian_def}
\end{equation}
where $\theta$ is a complex Gaussian random variable with unit variance.

Now, instead of generating a set of points representing galaxies from this
density field, we generate a continuous density field which is already
affected by window functions. In this way one can study the effect of
window functions without being affected by shot noise. This can be done
by inverse-Fourier transforming $\delta({\mathbf k})$ to obtain the real-space
density field, $\delta({\mathbf r})$, and multiplying $\delta({\mathbf r})$
by the window function, $W({\bf r})$. The power spectrum is estimated from
the Fourier transform of $\delta_g({\mathbf r})=\delta({\mathbf r})W({\bf r})$
as $\hat{P}_g({\bf k})=\frac{1}{W_{\rm sq}}|\delta_g({\bf k})|^2$.
One can show that the estimated power spectrum is given by
\begin{eqnarray}
 \langle\hat{P}_g({\bf k})\rangle&=&\frac{1}{W_{\rm sq}}
 \langle|\delta_g({\bf k})|^2\rangle
 =\frac{1}{W_{\rm sq}}\int\frac{d^3q_1}{(2\pi)^3}\int\frac{d^3q_2}{(2\pi)^3}~
 \langle\delta({\bf q}_1)\delta({\bf q}_2)\rangle
 W({\bf k}-{\bf q}_1)W(-{\bf k}-{\bf q}_2) \nonumber\\
 &=&\frac{1}{W_{\rm sq}}\int\frac{d^3q}{(2\pi)^3}~
 P({\bf q})|W({\bf k}-{\bf q})|^2 \ ,
\label{eq:expectation}
\end{eqnarray}
which agrees with eq.~\ref{eq:Pconv_def}, except for the shot noise term.

\begin{figure}[t]
\centering{
\includegraphics[width=1\textwidth]{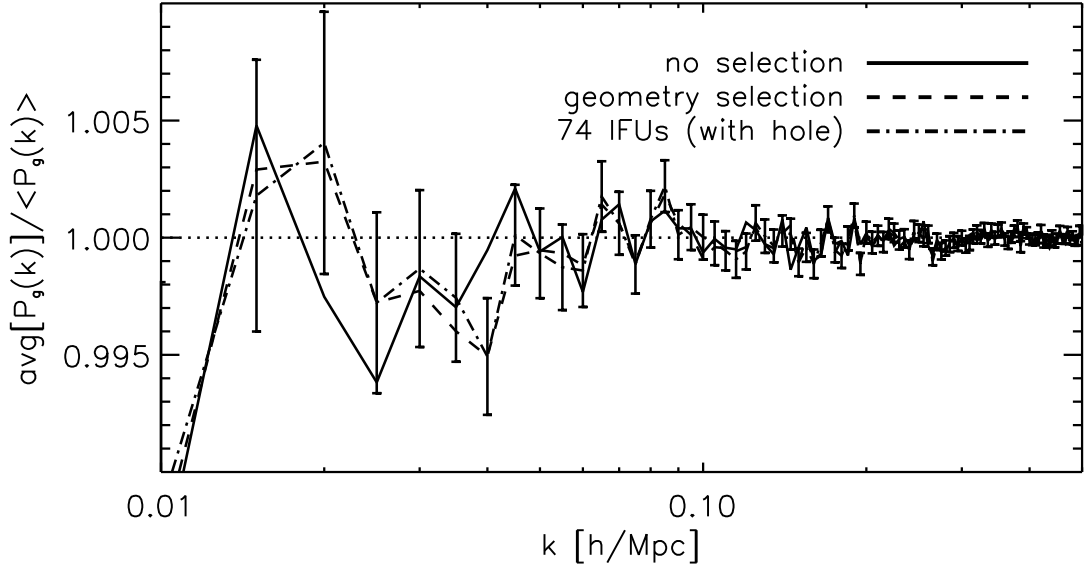}
}
\caption{
 Ratio of the average of 1000 power spectra measured from Gaussian
 realizations to the underlying power spectrum convolved with each of
 the window functions, i.e., we test validity of eq.~\ref{eq:expectation}.
 The solid, dashed, and dot-dashed lines are for no selection, the geometry
 selection, and the selection of 74 IFUs with a central hole, respectively.
 The error bars show the errors on the mean for 74 IFUs with a hole.
 The error bars for no- and geometry-selections are smaller.
}
\label{fig:gaussian_pk}
\end{figure}

To evaluate the performance of Gaussian realizations, we generated 1000
realizations for no selection, the geometry selection, and the selection
of 74 IFUs with a central hole. Figure~\ref{fig:gaussian_pk} displays the
ratio of the average of 1000 power spectra measured from Gaussian
realizations to the underlying power spectrum convolved with each of
the window functions. Fractional differences are less than 1\% for all
Fourier modes.

\bibliography{references}
\end{document}